\documentclass[10pt,journal,compsoc]{IEEEtran}
\usepackage{graphbox}
\usepackage{flushend}
\usepackage{algorithm}
\usepackage[noend]{algpseudocode}
\usepackage{amsfonts}
\usepackage{amsmath}
\usepackage[usenames, dvipsnames]{color}
\usepackage{ragged2e}
\usepackage{subfigure}
\usepackage{arydshln}

\usepackage[utf8]{inputenc}
\usepackage[T1]{fontenc}

\ifCLASSOPTIONcompsoc
  \usepackage[nocompress]{cite}
\else
  \usepackage{cite}
\fi
\ifCLASSINFOpdf
\else
\fi
\begin{document}
%
\title{Con-Pi: A Distributed Container-based Edge and Fog Computing Framework}
%
%
%
%

\author{Redowan Mahmud ~\IEEEmembership{Member,~IEEE}
        and Adel N. Toosi ~\IEEEmembership{Member,~IEEE}
\IEEEcompsocitemizethanks{\IEEEcompsocthanksitem R. Mahmud and A. N. Toosi are with the \textbf{Dis}tributed Systems and \textbf{Net}work Applications (\textbf{DisNet}) Laboratory, Department of Software Systems and Cybersecurity, Faculty of Information Systems, Monash University, Australia. \protect
\\E-mail: \{redowan.mahmud,adel.n.toosi\}@monash.edu}
\thanks{Manuscript received ~; revised ~.}}

%
%

\markboth{}%
{Mahmud \MakeLowercase{\textit{et al.}}: Con-Pi: Container-based Fog and Edge computing framework}
%



%
\IEEEtitleabstractindextext{%
\justify{\begin{abstract}
Edge and Fog computing paradigms overcome the limitations of cloud-centric execution for different latency-sensitive Internet of Things (IoT) applications by offering computing resources closer to the data sources. Small single-board computers (SBCs) like Raspberry Pis (RPis) are widely used as computing nodes in both paradigms. These devices are usually equipped with moderate speed processors and provide support for peripheral interfacing and networking, making them well-suited to deal with IoT-driven operations such as data sensing, analysis, and actuation. However, these small Edge devices are constrained in facilitating multi-tenancy and resource sharing.
The management of computing and peripheral resources through centralized entities further degrades their performance and service quality significantly. To address these issues, a fully distributed framework, named \textit{Con-Pi}, is proposed in this work to manage resources at the Edge or Fog environments. Con-Pi exploits the concept of containerization and harnesses Docker containers to run IoT applications as micro-services. 
The software system of the proposed framework also provides a scope to integrate different IoT applications, resource and energy management policies for Edge and Fog computing. Its performance is compared with the state-of-the-art frameworks through real-world experiments. The experimental results show that Con-Pi outperforms others in enhancing response time and managing energy usage and computing resources through its distributed offloading model. Further, we have developed an automated pest bird deterrent system using Con-Pi to demonstrate its suitability in developing practical solutions for various IoT-enabled use cases, including smart agriculture.  
\end{abstract}}

\begin{IEEEkeywords}
Edge computing, Raspberry Pi, Microservice, Container, Fog computing, Docker.  
\end{IEEEkeywords}}

\maketitle

\IEEEdisplaynontitleabstractindextext

%
\IEEEpeerreviewmaketitle

\IEEEraisesectionheading{\section{Introduction}}\label{sec_introduction} 

%
%
%
%
%
\IEEEPARstart{I}{}n recent years, the theory of Internet of Things (IoT) is being rapidly adopted to create different Cyber-Physical Systems (CPSs) for health care, industry automation, smart agriculture and city management~\cite{bakerRef}. The operations of these CPSs, such as data collection, analysis, event detection and actuation, are explicitly managed by a set of IoT applications. The real-time interactions of IoT applications with the CPSs get obstructed when they are hosted in Cloud datacentres for execution. Usually, Cloud datacenters reside at a multi-hop distance from the IoT-enabled CPSs~\cite{hopCount}. Cloud-based solutions not only degrade the Quality of Service (QoS) of the applications but also significantly affect the responsiveness~\cite{Realtime}. Edge and Fog paradigms solve this problem by extending the computing facilities closer to the CPSs. In Edge computing, the IoT applications are hosted on the gateways; whereas, in Fog computing, any device within the Local Area Network (LAN) or coverage area of Tier 2 and 3 Internet Service Providers (ISPs) are used for the application execution \cite{TSC-1}. Thus, they improve the service delivery time and responsiveness of the applications. They also reduce network congestion by resisting the transfer of a considerable amount of data to Cloud datacenters~\cite{NetworkSlice}.
\par Conceptually, Edge/Fog computing can exploit routers and switches to run the IoT applications~\cite{affinity}. As these equipment need to perform continuous networking operations, they are often considered ill-suited for executing complex IoT applications such as image processing and robotic navigation~\cite{Ritu-FGCS}. Different computing devices such as personal computers, smart phones, cloudlets, micro datacentres and nano servers can also be used as the elements of Edge and Fog computing environments. However, these devices provide fewer supports for interfacing external sensors and actuators. Additionally, their outdoor deployment is subjected to cost and energy-related issues. To address these shortcomings, the usage of ARM-based single-board computers (SBCs) like Arduino and Raspberry Pi (RPi) for Edge and Fog computing is becoming increasingly popular~\cite{piFogBed}.
\par Similar to fully functional computers, SBCs are equipped with processors, memory, storage, display and Ethernet ports, but less expensive~\cite{RPi}. Moreover, the General-purpose input/output (GPIO) pins of devices like RPis help interfacing with external sensors, power sources and actuators. However, SBCs are bare metals in nature and less supportive of multi-tenancy. Therefore, virtualization is necessary, although hypervisor-based virtualizations are not suitable for resource-constrained SBCs due to heavyweight management operations. In such cases, containerization is regarded as the viable solution \cite{Pahl}. 
\par Application containers, such as Docker, use the operating system-level virtualization to encapsulate application logics as microservices. They execute microservices by provisioning a minimal amount of computing resources~\cite{microservice}. Containers are lightweight and easily configurable compared to virtual machines. They facilitate secured application execution through explicit isolation. Nevertheless, the QoS-satisfied service delivery of IoT applications largely depends on how the containers are managed, especially when they are hosted in embedded machines like RPis \cite{TSC-3}.       
\par Recently, different frameworks for managing resources in Edge/Fog environments have been developed ~\cite{EiF}~\cite{FogBus}. They barely discuss the containerization of computing nodes and fail to deal with resource-overhead. Some other initiatives emphasize managing containers on SBCs~\cite{FogPlan}~\cite{IoTDoc}. However, they operate in a centralized manner which is often unscalable and not suitable to meet the unstable situations of networks \cite{afrinICSOC}. Moreover, they provide limited scope for sharing resources and integrating different policies to manage resources for application execution. In particular, existing frameworks have paid limited attention to the critical role of efficient energy management.

To overcome the limitations of existing frameworks, in this work, we propose a novel container-based solution, named \textit{Con-Pi} to harness the capabilities of SBCs for Edge/Fog computing. Con-Pi poses a service-oriented architecture that makes it highly scalable. Con-Pi facilitates autonomic initiation of containers on RPis and supports creating different topologies, including mesh and hierarchical. It functions in a fully distributed manner across multiple SBCs and assists in resource sharing. Con-Pi is also among the few that support harvesting renewable energy sources such as solar power and managing energy storage devices such as a battery for SBCs at the Edge and Fog. Additionally, the software system of Con-Pi is agile and lightweight because of the simplified synthesis of its component programming languages and libraries. It also offers seamless integration of various resource management and discovery policies for power-constrained Edge/Fog environments. The major contributions of this paper are:
\begin{itemize}
\item A fully-distributed framework for containerized SBCs to share resources in Edge and Fog computing environments while executing applications as microservices.
\item A software system where different applications and resource management policies, including energy management for Edge and Fog computing, can be seamlessly integrated. 
\item A a proof-of-concept real-world prototype developed using Con-Pi and a group of RPis with an application demonstrator in smart agriculture. 
\end{itemize}
\par The rest of the paper is organized as follows: Section \ref{sec_related} describes related frameworks. The hardware components, software system and implementation of Con-Pi are discussed in Section \ref{sec_framework}. Section \ref{sec_aspects} discusses the important features of the proposed framework. The performance of Con-Pi in respect to existing frameworks is evaluated in Section \ref{sec_performance}. As a case study, a pest bird repellent system developed using Con-Pi is presented in Section \ref{sec_case}. Section \ref{sec_future} denotes potential direction for further improvement of the proposed Con-Pi framework. Finally, Section \ref{sec_conclusions} concludes the paper.  
\section{Related work} \label{sec_related}
There exist different frameworks for managing applications and resources at the Edge and Fog infrastructure where the containerization of computing nodes is not enforced. For example, Tuli et al.~\cite{FogBus} developed \textit{FogBus} framework that integrates IoT devices, Fog nodes and Cloud infrastructure. It only follows the master-slave approach for task distributions. FogBus operates in bare metal and lacks support for containerization. To implement FogBus, a synthesis of multiple programming languages is required. An et al.~\cite{EiF} developed another framework named \textit{EiF}. It runs on the gateway level and incorporates artificial intelligence services. EiF also embeds a predictive network resource allocation mechanism that enhances the performance of network slices and ensures scalability during resource provisioning. However, it is constrained in integrating external policies. Nguyen et al.~\cite{Market-based} proposed a \textit{Market-based} Fog-based framework that allocates multiple resources to applications according to the budget constraints. It adopts a centralized solution to solve the resource allocation problem. The privacy-preserving algorithm associated with the framework handles the sensitive user data. Additionally, a multi-level DDoS mitigation framework named \textit{MLDMF} is developed by Yan et al.~\cite{MLDMF}. It targets the industrial IoT context. In MLDMF, the concept of Software-defined networking (SDN) is integrated. The framework operates across Edge, Fog and Cloud infrastructure. In another work, Ghosh et al.~\cite{Mobi-IoST} proposed the \textit{Mobi-IoST} framework for dealing with mobility issues in integrated Fog-Cloud environments. It models the Spatio-temporal data of moving agents for probabilistic decision making. \textit{Mobi-IoST} also manages the energy consumption of computing nodes and the QoS of applications. Nevertheless, Borthakur et al.~\cite{SmartFog} discussed a framework named \textit{SmartFog} to facilitate big data analysis in Fog environments. It adopts machine learning techniques to extract and analyse data from wearable IoT devices. SmartFog operates at the gateway level and maintains a seamless connection with Cloud datacentres. Similarly, Usman et al. developed an intelligent framework named \textit{RaSEC} for managing mobile data in Edge and Fog computing with reliability and security~\cite{RaSEC}. RaSEC operates in three phases ensuring lightweight aggregation, encryption and analysis of data. It also maintains an event-driven interaction with Cloud for detailed analysis depending on the moving objects in the data. However, this framework lacks support for virtualized instances in Edge/Fog computing.  
\begin{table}[!t]
\scriptsize
\centering 
\caption{A Summary of related work and their comparison}\label{Tab:summary} 
\begin{tabular}{|p{1.95 cm} p{0.85 cm} p{0.5 cm} p{0.95 cm} p{1.3 cm} p{0.85 cm}|}
 \hline
 Work & Enables \newline containers & Targets SBCs & Integrates multiple policies & Operates dispersedly & Shares \newline resources \\\hline
FogBus~\cite{FogBus} & &  & \checkmark  & & \checkmark \\
EiF~\cite{EiF} &  &  &  & \checkmark & \checkmark \\
Market-based~\cite{Market-based} &  &  & \checkmark &  & \checkmark \\
MLDMF~\cite{MLDMF} &  & & & \checkmark & \checkmark \\
Mobi-IoST~\cite{Mobi-IoST} &  & & \checkmark & \checkmark & \\
SmartFog~\cite{SmartFog} &   & \checkmark &  & \checkmark & \\
RaSEC~\cite{RaSEC}  &   &  & \checkmark & \checkmark & \\
Foggy~\cite{Foggy} & \checkmark &  & \checkmark & \checkmark &  \\
IoTDoc~\cite{IoTDoc} & \checkmark & \checkmark & \checkmark & &  \\
Ext. Kura~\cite{Kura} & \checkmark & \checkmark & & &  \\
Ad-Hoc Edge~\cite{Ad-Hoc} & \checkmark & \checkmark & \checkmark & &  \\
FogPlan~\cite{FogPlan} & \checkmark &  &  & \checkmark &  \\
OpenStack~\cite{OpenStack} & \checkmark  & & \checkmark & \checkmark & \\
\textbf{Con-Pi} & \checkmark  & \checkmark  & \checkmark & \checkmark  & \checkmark   \\\hline
\end{tabular}  
\end{table}                       
\par On the contrary, there are some other frameworks where containerization is considered an integral part. For example, the \textit{Foggy} framework~\cite{Foggy} proposed by Yigitoglu et al. supports dynamic resource provisioning and automatic application placement in Fog computing environments. The container-based virtualization adopted in the Foggy framework helps system isolation and overhead management. Noor et al.~\cite{IoTDoc} proposed another framework named \textit{IoTDoc} to initiate and operate containers in the IoT devices. It also facilities user access to these containers through Cloud services. In IoTDoc, the interactions between users and IoT devices are managed by a centralized controller. In~\cite{Kura}, Bellavista et al. extended the open-source \textit{Kura} framework by introducing the concept of the local broker during IoT and Cloud service integration. It adopts a centralized approach. By enabling mesh topology among IoT gateways, the extended Kura framework scales up the computing infrastructure. In another work, Ferrer et al.~\cite{Ad-Hoc} proposed \textit{Ad-Hoc Edge} framework for creating Edge computing infrastructure dynamically. In Ad-Hoc Edge, an orchestrator coordinates the deployment of applications on the computing nodes. A logical registry is also managed in Ad-Hoc Edge to track the available Edge devices in the system. Additionally, Yousefpour et al.~\cite{FogPlan} discussed the \textit{FogPlan} framework that dynamically deploys and releases applications on Fog infrastructure based on the latency constraints and resource availability. It focuses on application-centric QoS requirements and incorporates two greedy algorithms to minimize overall application execution cost. Moreover, Merlino et al.~\cite{OpenStack} developed an \textit{OpenStack}-based framework to discover available containers at the Edge, Fog and Cloud infrastructure. It incorporates horizontal and vertical offloading patterns for engineering the workloads.
\par Table \ref{Tab:summary} provides a summary of existing frameworks. These frameworks do not facilitate resource sharing among multiple RPis through containers. They are often operated by a centralized entity and provides limited support to integrate customized policies for energy management and resource discovery. The proposed Con-Pi overcomes these limitations and offers Application Programming Interfaces (APIs) that assist users and developers in interacting with the computing infrastructure and peripheral devices in a simplified and lightweight manner.

\section{Con-Pi Framework} \label{sec_framework} 
\begin{figure}[!t]
\centering 
\includegraphics[width=88mm, height= 55mm]{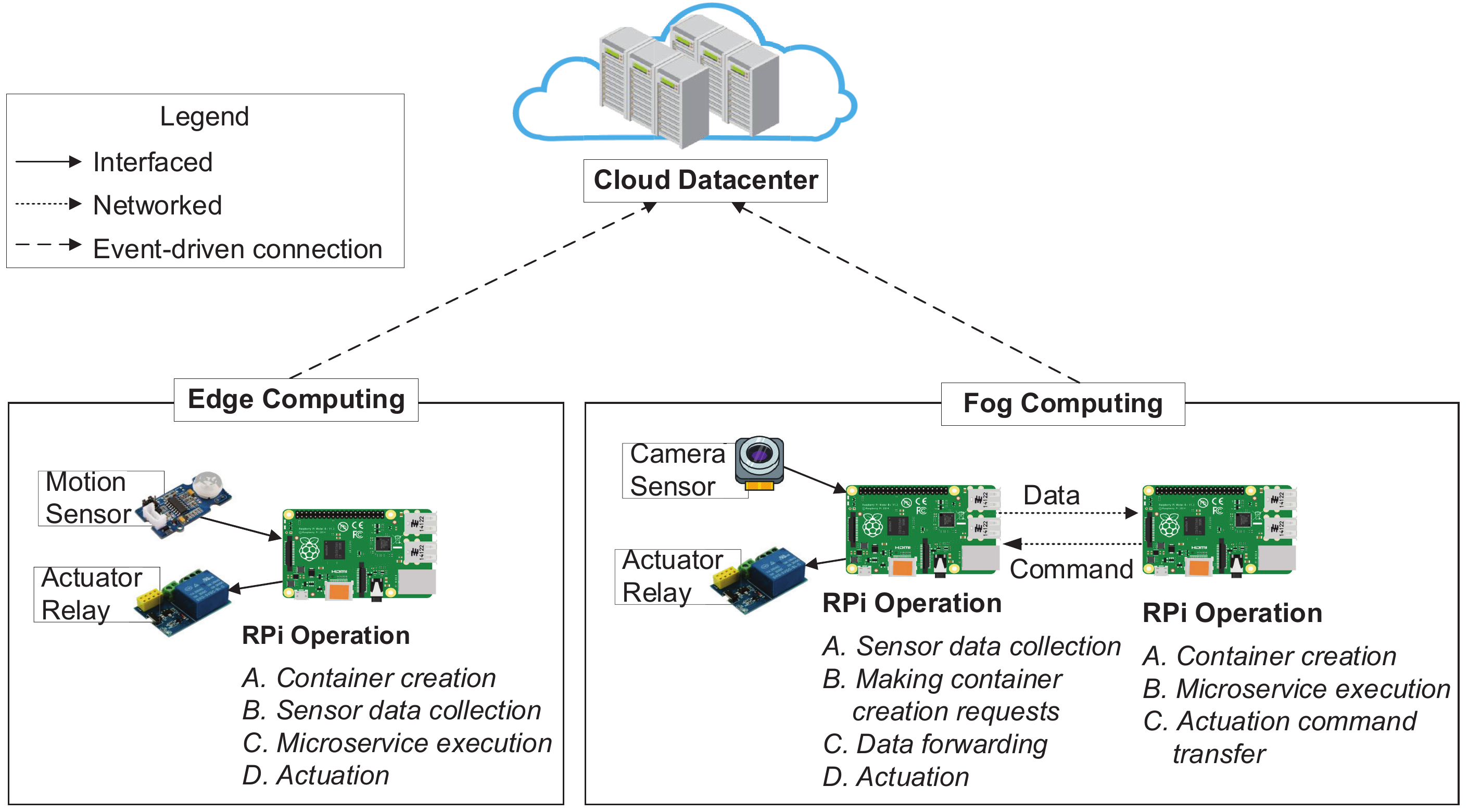}
\caption{Con-Pi-enabled Computing Environments}
\label{Fig:Con-PiBasic}
\end{figure}

A high-level view of Edge/Fog environments supported by Con-Pi is depicted in Fig. \ref{Fig:Con-PiBasic}. The detailed description of Con-Pi framework is given in the following subsections.   
\subsection{Hardware Element}
Our proposed framework, Con-Pi, is designed to be working with single-board computers (SBCs) such as Raspberry Pis (RPis). In the rest of the paper, the implementation of Con-Pi is discussed on RPis which is used in our prototype system. Con-Pi deals with various hardware, including RPi, PiJuice HATs, sensors and actuators, as discussed below. 
\par \textit{\textbf{RPi}}: 
SBCs, in particular RPis, offer run-time environments for most of the programming languages, including Java and Python. Most importantly, each RPi has a set of GPIO pins that assists interfacing with various peripheral devices, including sensors and actuators. RPis can be run using both alternative and direct power sources. Con-Pi enables an RPi to run multiple containers independently and can integrate different RPis to work collaboratively.     
\par \textit{\textbf{PiJuice HATs}}: Unlike desktop computers, SBCs are highly feasible to deploy at the outdoors and remote places. In such cases, it often becomes difficult to arrange the grid-based electricity to run them. PiJuice HAT is a portable power platform that supplies uninterrupted energy to run RPis using a rechargeable battery. PiJuice HAT uses five GPIO pins to work and provides APIs to check energy levels, set wake-up time, and manage downtimes. Con-Pi monitors the battery charge of RPis through PiJuice HAT while making decisions on container and data management.
\par \textit{\textbf{Sensors and Modules}}: Different temperature, humidity, motion, infrared, touch screen, navigation, and camera modules can be directly connected with the RPi. They help RPis in perceiving the external world. Con-Pi provides a scope to the application developers for interpreting the data generated by various sensors and modules. It also assists in forwarding sensor and module data among the RPis.    
\par \textit{\textbf{Actuators}}: SBCs can be coupled with a variety of actuators to initiate actuation based on the outcomes of data processing. It can also be incorporated with third-party devices such as sound systems, portable displays and laser modules through relay switches. Con-Pi supports the transferring of actuation commands across multiple RPis so that the physical actions can be triggered remotely from the place of data processing. Moreover, the actuators can use their power source to run or depend on the RPis.
\subsection{Software System}
\begin{figure}[!t]
\centering 
\includegraphics[width=88mm, height= 75mm]{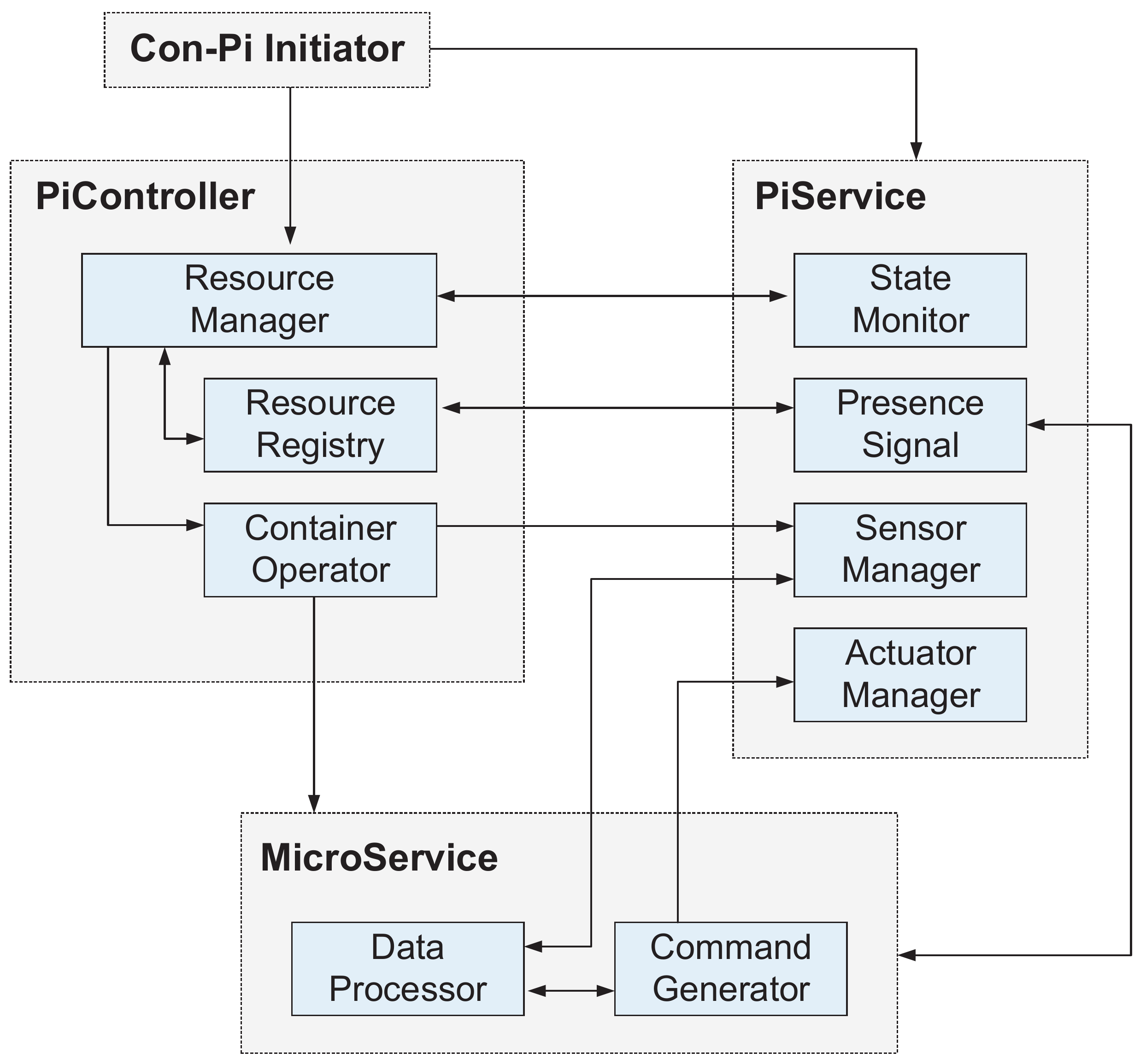}
\caption{Software components within the Con-Pi framework}
\label{Fig:Con-PiSoft}
\end{figure}

Since the Con-Pi framework operates in a distributed manner, its lightweight software system should be installed on each RPi. The Con-Pi software system consists of \textit{PiController} and \textit{PiService} as shown in Fig. \ref{Fig:Con-PiSoft}. Both assist the execution of microservices in RPis. \textit{Resource Manager}, \textit{Resource Registry} and \textit{Container Operator} work as the functional blocks for PiController. On the other hand, PiService is composed of \textit{State Monitor}, \textit{Presence Signal}, \textit{Sensor} and \textit{Actuator Manager}. These software components are explicitly modularized so that they can be easily linked with each other using Hypertext Transfer Protocol (HTTP) or Representational State Transfer (RESTful) APIs even when they are hosted on different RPis. There is another software component named \textit{Con-Pi Initiator} that starts the operations of PiController and PiService simultaneously within an RPi. The invocation of Con-Pi Initiator can be programmed to be event-driven or automatic. The Con-Pi software system is discussed in the following subsections.    
\subsubsection{PiController}         
\hspace*{12pt} \textbf{Resource Manager}: This component incorporates a set of algorithms and policies that manage the computing resources within an RPi. While making resource management decisions, the contextual information of the RPi is extended from the State Monitor of PiService to the Resource Manager. It also interacts with the Resource Registry to get the references of other RPis for resource and data sharing. Based on the outcome of associated algorithms and policies, the Resource Manager initiates commands for Container Operator to control the execution of microservices.      
\par \textbf{Resource Registry}: It executes a resource discovery algorithm that identifies all accessible RPis from the host RPi. Resource Registry consistently updates the references of RPis based on the Presence Signal information from the respective PiService. Resource Registry also maintains a data structure to track the past performance of other RPis. It assists the Resource Manager in determining the suitable RPi to request for resource sharing. 
\par \textbf{Container Operator}: The creation and termination of containers are handled by this component. Container Operator injects necessary arguments such as the reference of microservice and the address of RPi hosting the PiService to the containers through APIs. It also resists the Sensor Manager to forward data streams for recently terminated containers and updates their executor information at the Resource Registry.   
\subsubsection{PiService}
\hspace*{12pt} \textbf{State Monitor}: This component extracts the contextual information such as resource configuration, network bandwidth, residual battery lifetime, location, current memory, storage and CPU usage from the host RPis. Based on the requirements of the Resource Manager, State Monitor forwards such information to PiController and participates in making the resource management decisions.
\par \textbf{Presence Signal}: This software component periodically generates heartbeat signals for the Resource Registry of PiController to notify the presence of the host RPi. Presence Signal also informs the status of microservice execution to the respective PiController for tracking the run-time performance of the RPis. It helps in detecting the sudden failure of RPis. 
\par \textbf{Sensor Manager}: For each sensor and module connected with the host RPi, this component provides application programming interfaces to support the collection of generated data. Moreover, Sensor Manager assists in tuning the sensing frequency and data configuration of the sensors and modules as per the requirements of the microservices. It can start forwarding data to the microservice soon after the initiation of the respective container or can wait until the microservice interacts.
The connected sensors work in \textit{pull}/\textit{push} modes. In the pull mode, microservices get sensor data from Sensor Manager only after sending requests, while in the push mode, the Sensor Manager autonomously sends sensor data to the microservices. The Sensor Manager's push mode is designed to be working based on the  \textit{publish}/ \textit{subscribe} pattern and maintains a data structure that helps microservices to subscribe and unsubscribe from any sensor data stream during run-time.  
\par \textbf{Actuator Manager}: This component exclusively interacts with the microservice for receiving the actuation command based on the outcome of data processing operations. Actuator Manager can support the simultaneous initiation of multiple actuators with customized outputs.      
\par However, in order to align with the PiController and PiService, microservices require to meet a specific architecture. The main aspects of this architecture are listed as follows. 
\begin{itemize}
\item The microservices should have a scope to parse the instructions forwarded from the Container Operator. 
\item Software components within a microservice such as Data Processor and Command Generator require modularization so that they can be configured separately without affecting the peers' functions.  
\item The Data Processor component of the microservice needs to maintain a multi-dimensional interaction with the Sensor Manager so that it can receive or pull the sensor-generated data alternatively.
\end{itemize}   
\begin{figure*}[!t]
\centering 
\includegraphics[width=0.85\textwidth]{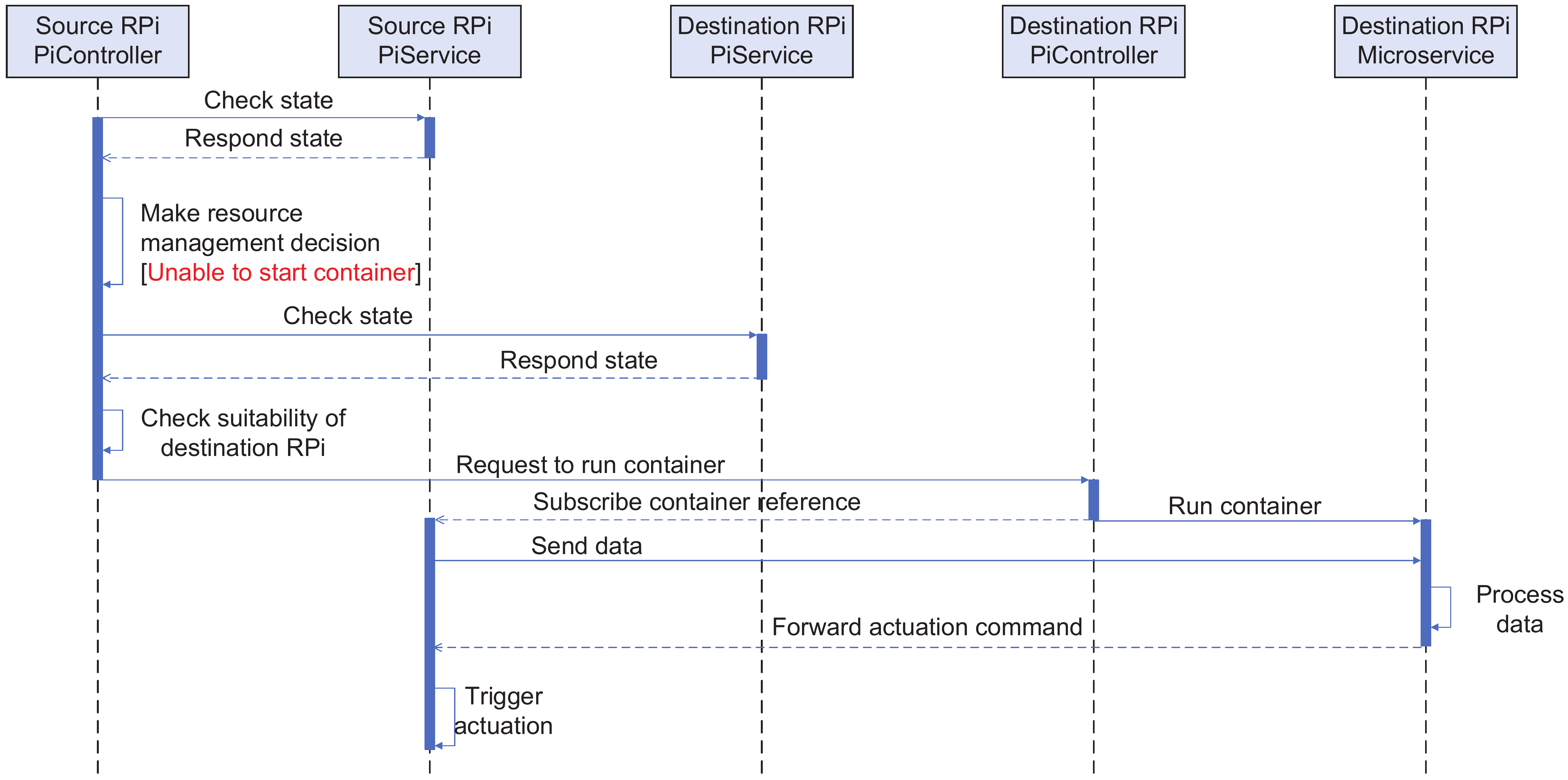}
\caption{Interactions during resource sharing}
\label{Fig:Con-PiInt}
\end{figure*}
\subsection{Implementation} 
The execution environments, programming languages, software packages and APIs used in implementing Con-Pi framework are listed below.  
\par \textbf{Operating System}: Con-Pi framework prefers Raspbian Buster and Raspbian Stretch as the operating system for the RPIs. Both operating systems are the updated versions of Debian Linux that facilitate more than 35,000 software packages to optimize the performance of RPis \cite{Raspbian}. 
\par \textbf{Programming Language}: Python 3 has been used to develop the software components for PiController and PiService because of its cross-platform assistance \cite{python}. The APIs for extracting data of PiJuice HATs are also designed in Python 3. However, the microservices managed by the Con-Pi framework can be developed in any programming language, including Python 2 and Python 3.
\par \textbf{Communication}: Con-Pi follows a service-oriented architecture. Each software component within PiController and PiService are modelled as a Flask-based web service. Flask is a lightweight web framework developed in Python programming language that does not have any database abstraction layer and third-party dependencies. Con-Pi uses HTTP-based RESTful APIs to establish the network among the RPIs and transfer data between different software components, which is faster than the contemporary techniques such as shared database and message passing \cite{Rest}. 
\par \textbf{Container}: Con-Pi framework uses Docker platform for containerization of applications and its logics. Docker facilitates easy packaging and management of microservices within containers. To ensure reproducibility of containers on multiple RPis, the specifications are stored in a Dockerfile \cite{Docker}. The container images built from the same Dockerfile offers identical functions. Moreover, the compiled Docker container images can be distributed at a large scale through Docker Hub. Docker also provides Python-based APIs that help creation, setting of environment variables and deployment of Docker containers programmatically. The initiation of Docker containers is as fast as running a normal application process. Furthermore, Docker permits to commit the changes of an operational container to another Docker image with an embedded version control mechanism. These features of Docker assist Con-Pi to customize the execution of microservices on the RPis. 
\par \textbf{Auto-start}: To schedule the auto-start of Con-Pi framework during the boot-up of RPis, the Crontab utility is used. Crontab facilitates the execution of a program on RPis with privileged rights. Therefore, its response time is faster. Corntab is operated through a config file that requires to be edited during the installation of Con-Pi on the RPis. 
\par \textbf{Installation}: From the implementation perspective, the Con-Pi framework requires support from multiple third-party software packages. Individual installation of these packages is tedious. Therefore, Con-Pi framework incorporates a bash executable script that makes the RPis compatible to host the framework. This script installs the Docker platform and sets the right for privileged execution. Additionally, it augments the Python-based libraries for Flask, Docker, PiCamera, GPIO, PiJuice and Representational state transfer to the run-time environments of RPis and updates Corntab for the auto-start of Con-Pi software components.    
\section{Features of Con-Pi} \label{sec_aspects}
In comparison to the existing frameworks, Con-Pi possesses several important features that indicate its novelty. For example, Con-Pi facilitates simultaneous computing and peripheral resource sharing and integration of multiple resource management and discovery policies, which are missing in most contemporary solutions. Due to its fully distributed software components, Con-Pi also supports the creation of different topologies within an Edge/Fog environment that ultimately meets the requirements for dynamic scalability. A detailed description of these features is given in the following subsections.
\subsection{Resource Sharing}
Fig.~\ref{Fig:Con-PiInt} depicts the interactions among multiple RPis while sharing resources through Con-Pi. At the booting time, the PiController of source RPi initiates this operation by making a request to the corresponding PiService for checking the state of the RPi. Con-Pi provides a scope to check various states of the RPis, including battery lifetime and CPU usage, simultaneously as per the requirements of the resource management policies. Given the perceived state, if the PiController identifies the source RPi incapable of starting the containers, it communicates with the PiServices of other RPis for analyzing their states to select a destination RPi for extending resources. In this case, if an RPi is already energy drained or overloaded either with its containers or offloaded containers, or both, that RPi will not be presented to the source RPi as a prospective resource sharing destination by Con-Pi for avoiding further conflicts. Additionally, while selecting the suitable RPi from the prospective destinations, the resource management policy embedded at the PiController of source RPi can explicitly investigate their network signal strength, the residual battery lifetime and the current resource utilization. After the selection, the PiController of source RPi sends a request to the destination RPi to run a container. This request also includes the address of the source RPi PiService and the reference of the microservice. Later, the destination RPi's PiController triggers the execution of the microservice by launching a container. Concurrently, the PiController as a mediator conducts the handshaking between the container and the source RPi PiService through reference subscribing so that both can interact with each other without the interventions of other Con-Pi components. During such interactions, the source RPi PiService sends sensor data to the microservice at destination RPi. After processing data, the microservice at the destination RPi forwards the actuation command to the source RPi PiService. In other words, the execution of microservices and their respective containers are offloaded among RPis while the sensor data and actuation interfaces can be accessed remotely. 
\par Similar to Con-Pi, the EiF framework~\cite{EiF} facilitates resource sharing among multiple computing entities. However, Con-Pi is unconventional as it not only helps in sharing computing resources but also simplifies the access to the peripheral sensor and actuator devices through PiService. Con-Pi also supports resource sharing in a scalable manner, which means a source RPi can extend resources from different destination RPis to run multiple containers concurrently. Furthermore, Con-Pi facilitates input and command-dependent interactions between a source RPi and a destination RPi. Since a small amount of data and short messages are exchanged during these interactions, any varying network conditions have a very minimal effect on the resource sharing performance of Con-Pi.   
\begin{algorithm}[!t]
\footnotesize
\caption{Energy-aware Resource Management}\label{algo-share}
\begin{algorithmic}[1]
\Procedure{ManageResource}{}
\State $ \pi^s.Local \gets  false$
\State $ \pi^s.Remote \gets  false$
\State $ \pi^s.SelectedRPi \gets  null$
\While{true}
\State $ \pi^s.Battery \gets getBattery(\pi^s.serviceAddress)$
\If{$\pi^s.Battery > \alpha$}
\If{$\pi^s.Local = null \text{ \& } \pi^s.Remote = null $}
\State $dockerStart(\pi^s.controlAddress \newline \hspace*{3.3 cm},\pi^s.serviceAddress, dockerImage)$
\State $\pi^s.Local \gets  true$
\Else {$\hspace*{0.1 cm} \pi^s.Local = null \text{ \& } \pi^s.Remote \neq null \newline \hspace*{1.8 cm} \text{ \& }  \pi^s.Battery > \beta$}
\State $dockerStart(\pi^s.controlAddress \newline \hspace*{3.3 cm},\pi^s.serviceAddress, dockerImage)$
\State $\pi^s.Local \gets  true$
\State $dockerStop(\pi^s.SelectedRPi.controlAddress \newline \hspace*{3.3 cm
}, \pi^s.serviceAddress, dockerImage)$
\State $\pi^s.Remote \gets  false$
\EndIf
\Else
\If{$\pi^s.Remote = null$}
\State $\Pi \gets \textsc{DiscoverRPis}(\pi^s)$
\For{$ \pi^d := \Pi$}
\State $ \pi^d.Battery \gets getBattery(\pi^d.serviceAddress)$
\If{$\pi^d.Battery > \gamma$}
\State $ \pi^s.SelectedRPi \gets  \pi^d $
\State $ break $
\EndIf
\EndFor
\State $dockerStart(\pi^s.SelectedRPi.controlAddress \newline \hspace*{3.3 cm
}, \pi^s.serviceAddress, dockerImage)$
\State $\pi^s.Remote \gets  ture$
\State $dockerStop(\pi^s.controlAddress \newline \hspace*{3.3 cm},\pi^s.serviceAddress, dockerImage)$
\State $\pi^s.Local \gets  false$
\EndIf
\EndIf
\If{$\pi^s.Remote = true$}
\State $ \eta = checkRemoteExecution(\pi^s.serviceAddress) $
\If{$\eta = false$}
\State $\pi^s.Remote \gets  false$
\EndIf
\EndIf
\EndWhile
\EndProcedure
\end{algorithmic}
\end{algorithm}
%
\subsection{Policy Integration}\label{Sec:policy}
The Con-Pi framework provides a broader scope for customized policy integration so that the computing and peripheral resources within the RPis can be managed and discovered in various ways. Such policies extend different APIs and run-time parameters of PiController to operate promptly and perform data offloading, resource sharing and load-balancing simultaneously. Con-Pi also models the PiController and PiService as separate web services. Hence, for any RPi, the resource management and discovery policies specify the addresses of PiController and PiService using the combination of \textit{$<$IP address, Port number$>$}. 
Two of such policies integrated into Con-Pi are discussed below. Here, syntactic notations of object-oriented programming languages are used for simplified understanding. 
\subsubsection{Resource management policy} \label{RMS}
Energy efficiency is considered as one of the driving factors for resource management in Edge/Fog environments. To model an energy-aware resource management scenario, we consider an Edge/Fog computing environment with a set of $n$ RPis $\Pi=\{\pi_1, \pi_2,\pi_3,...,\pi_n\}$ that requires to execute $m$ microservices over a period of $T$ timeslots. For the sake of simplicity, we assume microservices are homogeneous. At any timeslot $t=\{0, 1, 2, ...., T\}$, $\mu_{i,t}$ denotes the number of microservices executing on a RPi $\pi_i$. If RPi $\pi_i$ is the initiator (source) for $\upsilon_{i,t}$ number of microservices executing on $\pi_i$ at timeslot $t$, $(\mu_{i,t}-\upsilon_{i,t})$ defines the number of offloaded microservices to $\pi_i$. Furthermore, in such an Edge/Fog environment, RPis receive energy from solar power and charge their batteries to remain active. For any RPi $\pi_i$, the predicted solar power input and the residual battery charge at timeslot $t$ are denoted by $\kappa_{i,t}$ and $\delta_{i,t}$, respectively. Whether the RPi will continue its function during the timeslot, $t$ is determined by a binary variable $\chi_{i,t}$ as noted in Eq. \ref{Eq.chi}   
\begin{equation}\label{Eq.chi}
 \chi_{i,t}=
    \begin{cases}
      0, & \text{if}\ \kappa_{i,t} + \delta_{i,t-1} - \vartheta_{i,t} \leq 0 \\
      1, & \text{otherwise}
    \end{cases}
\end{equation}
Here, $\vartheta_{i,t}$ refers to the total amount of energy consumed by the RPi $\pi_i$ while executing $\mu_{i,t}$ microservices in timeslot $t$. If $\phi_i$ and $\varphi_i$ denote the per timeslot energy requirement of a microservice for computation and networking operations respectively, since there is no networking operations for locally initiated microservices, $\vartheta_{i,t}$ is calculated as:          
\begin{equation}\label{Eq.vartheta}
\vartheta_{i,t} = \phi_i \times \upsilon_{i,t} + (\phi_i + \varphi_i) \times (\mu_{i,t} - \upsilon_{i,t})
\end{equation}
However, depending on $\chi_{i,t}$, the slot-wise operative time $\tau_i$ of a RPi is calculated using Eq. \ref{Eq.tau}                
\begin{equation}\label{Eq.tau}
\tau_i = \sum_{t=1}^{t=T} \chi_{i,t}
\end{equation}
For the stability of such an Edge/Fog environment, it is expected to keep all RPis operational throughout the $T$ timeslots, which cannot be possible due to variations in the energy availability of the RPis. Hence, the proposed model aims to maximise the operational time of all RPis (having the least number of non-operative timeslots). To achieve this, the objective function of the proposed model aims to maximize the number of operative timeslots for an RPi with the minimum number of operative timeslots as shown in Eq. \ref{Eq.obj}. Constraints of this ILP problem ensure that the number of microservices is always equal to the number of active nodes (Eq. \ref{Eq.cons1}), and the sum of predicted solar power input and residual battery charge for any RPi $\pi_i$ does not surpass the maximum battery capacity $\zeta_i$ (Eq. \ref{Eq.cons2}). 
\begin{equation}\label{Eq.obj}
\max \min(\tau_i); \forall i \in \Pi, \forall t \in \{1,2,3, ..., T\}.
\end{equation}
subject to,\\
\begin{equation} \label{Eq.cons1}
\sum_{i=1}^n \mu_{i,t} = \sum_{i=1}^n \chi_{i,t}; \forall t \in \{1,2,3, ..., T\}.
\end{equation}
\begin{equation} \label{Eq.cons2}
\kappa_{i,t} + \delta_{i,t-1} \leq \zeta_i; \forall i \in \Pi, \forall t \in \{1,2,3, ..., T\}.  
\end{equation}
Considering the centralized nature of the proposed optimization model, the real-time requirements of managing microservices, and the computational and energy cost of solving the optimisation problem (Eq.~\ref{Eq.obj}) using ILP solvers (e.g., SCIP~\cite{solver}), it is not practical to apply this model in practice.
To address this challenge, Con-Pi encapsulates a sample greedy energy-aware resource management policy in the form of \textit{\textsc{ManageResource}} procedure (Algorithm \ref{algo-share}) for the modelled Edge/Fog environment.

\par For a source RPi $\pi^s$, Algorithm \ref{algo-share} at first initializes the run-time parameters such as $\pi^s.Local$, $\pi^s.Remote$ and $\pi^s.SelectedRPi$ to track the status of local and remote microservice executions and the reference of the selected destination RPi for resource sharing respectively (line 2-4). The subsequent operations of Algorithm \ref{algo-share} can be consolidated into three phases as listed follows. 
\par $\bullet$ \textit{Local microservice execution}: In this phase, the procedure starts and continues perceiving the battery lifetime percentage $\pi^s.Battery$ of source $\pi^s$  using the $getBattery$ API that takes the address of source RPi PiService $\pi^s.serviceAddress$ as argument (line 5-6). If $\pi^s.Battery$ is greater than threshold value $\alpha$ and other relevant parameters such as $\pi^s.Local$ and $\pi^s.Remote$ are unset, the policy invokes the \textit{dockerStart} API with the PiController and PiService addresses of the source RPi and the Docker image name, and initiates the microservice execution locally (line 7-10). 
\par $\bullet$ \textit{Transferring of remote microservice execution to local}: During the operations of a resource management policy, there might be some scenarios when the microservice execution is required to bring back to the source RPi from any previously selected destination RPi. According to Algorithm \ref{algo-share}, such operation is triggered when $\pi^s.Battery$ surpasses a certain threshold $\beta$ (line 11). In this case, the \textit{dockerStart} API is called with the same arguments as mentioned above (line 12-13). Additionally, the policy stops the remote execution of microservice at the selected destination RPi by passing the references of $\pi^s.SelectedRPi$ PiController, $\pi^s$ PiService and Docker image to the \textit{dockerStop} API (line 14-15). 
\par \par $\bullet$ \textit{Remote microservice execution}: Consequently, when the  $\pi^s.Battery$ of source RPi $\pi^s$ becomes lower than the threshold $\alpha$, the \textsc{ManageResource} procedure starts investigating the battery state of charge of other RPis for determining $\pi^s.SelectedRPi$ to offload the data and extend the resources (line 16-17). The abstract $discoverRPis$ procedure of PiController assists the resource management policy in performing this operation (line 18). According to the procedure, if the $\pi^d.Battery$ exceeds a given threshold $\gamma$ for any RPi $\pi^d$, its first-fit selection is made as $\pi^s.SelectedRPi$ (line 19-23). Later, the \textit{dockerStart} API is invoked with the references of $\pi^s.SelectedRPi$ PiController, $\pi^s$ PiService and Docker image to initiate the microservice execution remotely (line 24-25). At the same time, the source RPi terminates the local execution of the microservice (line 26-27). 
\par Algorithm \ref{algo-share} also facilitate the source RPi $\pi^s$ to seamlessly monitor the remote execution of the microservice using the $checkRemoteExecution$ API that takes the $\pi^s.serviceAddress$ as an argument and tune the related parameters accordingly (line 28-31). Thus, Con-Pi manages the computing and peripheral resources of RPis as per the energy available on their batteries that ultimately helps to increase the operational time of all RPis within the system.
\begin{algorithm}[!t]
\footnotesize
\caption{Logical Address-based Resource Discovery}\label{algo-disco}
\begin{algorithmic}[1]
\Procedure{DiscoverRPis}{$\pi^s$}
\State $ \Pi \gets null$
\State $ \pi^s.NetMask \gets getNetMask(\pi^s.NetInterface)$
\State $ \pi^s.Subnet \gets getSubnet(\pi^s.IP, \pi^s.NetMask)$
\State $ P \gets getAllIP(\pi^s.Subnet)$
\For{$ \pi^d := P$}
\State $ \eta \gets getPresence(\pi^d.serviceAddress)$
\If{$\eta = true$}
\State $ \Pi \gets \Pi \cup \pi^d $
\EndIf
\EndFor
\State $ return \text{ } \Pi $
\EndProcedure
\end{algorithmic}
\end{algorithm}
\subsubsection{Resource discovery policy}
Algorithm \ref{algo-disco} illustrates a sample resource discovery policy in the form of \textit{\textsc{DiscoverRPis}} procedure which can be customized or augmented with any resource management policies as per the intention of the user or system operator. Here, the procedure takes the reference of a source RPi $\pi^s$ as an argument and identifies the destination RPis based on the similarity of the logical address to $\pi^s$. At first, it sets necessary system parameters and identifies the \textit{Network Mask} for the communication interface of the source RPI $\pi^s$ using the $getNetMask$ API (line 2-3). Moreover, the $getSubnet$ API takes the IP address and Network Mask of $\pi^s$ as the arguments and helps in determining the associated \textit{Subnet} (line 4). Later, the IP addresses of all devices residing at the Subnet are accumulated using $getAllIP$ API (line 5). Finally, the devices are explored individually, and the presence of destination RPis are identified through the $getPresence$ API. In this operation, the PiService address of the RPis is explicitly used (line 6-9).     
\par Algorithm \ref{algo-share} and \ref{algo-disco} illustratively present relevant APIs and run-time parameters provided in the PiController of Con-Pi framework. Similar to Con-Pi, there exist other frameworks, including Foggy~\cite{Foggy} that facilitate policy integration for resource management. However, the policy tuning parameters supported by these frameworks are limited. Comparatively, Con-Pi endorses the coexistence of multiple policy tuning parameters such as latency sensitivity of the applications, network connectivity, data sensing frequency, computing capacity and battery state of charge of the RPis. These features make the Con-Pi novel and compatible to deal with the dynamics of Edge/Fog computing environments as per the service requirements.
\begin{figure*}[!t]
\centering 
\includegraphics[width=182mm, height= 40mm]{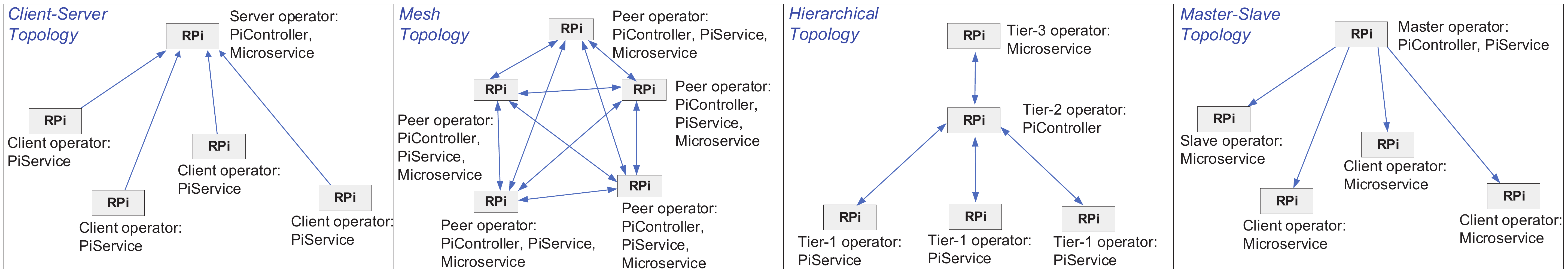}
\caption{Different topologies supported by Con-Pi}
\label{Fig:Topo}
\end{figure*}
%
\subsection{Support for Different Topologies}
The modular structure and the event-driven communication assistance make Con-Pi flexible to support different topologies simultaneously. Fig. \ref{Fig:Topo} depicts the topologies that can be modelled using Con-Pi. The distribution of the Con-Pi software system can differ from one topology to another. For example, in client-server topology, the server RPis host the PiController and microservice. The client RPis accommodate the PiService so that the clients can send data for processing to the servers. Conversely, in master-slave topology, the master RPis hold the PiController and PiService, and distribute data to the slave RPis that run the containers. However, the mesh topology is inherently endorsed by the Con-Pi framework where the PiController, PiService and microservices are hosted in peer RPis in an integrated manner. This default arrangement of Con-Pi software components is favourable to create dynamic clusters of RPis. By tuning the number of active communication links among the peer RPis through corresponding PiControllers, such clusters can also be made fully or sparsely connected, ultimately simplifying their scaling per the network and service requirements. Nevertheless, using Con-Pi, one of the possible configurations of hierarchical topology can be attained by placing the PiService of Con-Pi at the tier-1 (lower infrastructure level) RPis, while the PiController and microservice can be deployed at the tier-2 and tier-3 RPis, respectively. In this configuration, data are generated by the tier-1 RPis, and tier-2 RPis control the data flow towards the tier-3 RPis where the actual data processing happens. 
\begin{figure}[!t]
\centering 
\includegraphics[width=90mm, height= 40mm]{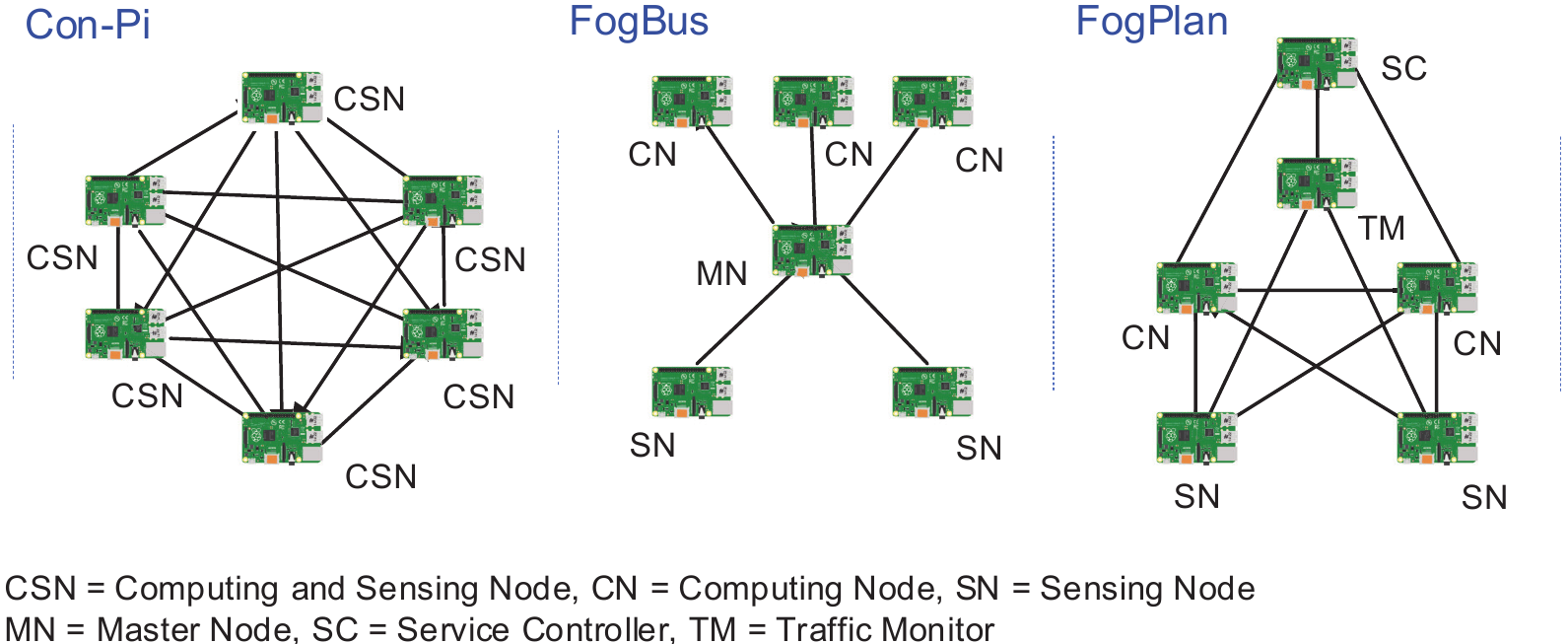}
\caption{Setup for Con-Pi, FogBus and FogPlan framework}
\label{Fig:Frame}
\end{figure}
\par The aforementioned illustrations are just the samples. The client-server, master-slave, mesh and hierarchical topologies can be modelled in diverse ways using Con-Pi. This feature makes Con-Pi more innovative than other frameworks, including FogBus, where only the master-slave topology can be modelled.
\subsection{Distributed and Scalable}
Each Con-Pi enabled RPi can autonomously run containers and terminate them, discover other RPis, and trigger remote execution without the intervention of any supervisor entity. Con-Pi also provides transparency during the local and remote operation of containers. Although Con-Pi is specialized in dealing with RPis, its software system is easily extensible, and due to using cross-platform programming language and services, Con-Pi can support heterogeneous devices simultaneously. In this case, the PiService is only required to customize according to the specification of the devices. These features make the Con-Pi highly distributed. 
\par Additionally, the Con-Pi framework enables the RPis not only to use their physical and peripheral resources effectively but also facilitates extending the resources of other peer RPis during uneven increment in load and sudden capacity failure. It also helps Con-Pi in attaining load scalability and fault tolerance. In addition, Con-Pi supports the encapsulation of applications as microservices hosted in Docker containers. Microservices inherently consume less physical resources than other virtualization techniques that promote multiple application execution in parallel, even in a resource-constrained RPi. Thus, Con-Pi assists space scalability. Moreover, the PiController and PiService of any Con-Pi-enabled RPi can be accessed through the Internet using standard protocols. Since the software system of Con-Pi is independent of each other, it is easier to expand and consolidate the number of RPis using Con-Pi dynamically. Consequently, it boosts scalability. 
\par Similar to Con-Pi, the FogPlan framework~\cite{FogPlan} operates in a distributed and scalable manner. However, FogPlan only distributes the computing responsibilities to the Fog nodes, whereas the controlling operations such as resource management and data offloading are maintained through a centralized entity. In this context, Con-Pi outperforms others as it ensures explicit distribution of control and computation operations among the Edge/Fog nodes. 
\section{Performance Evaluation} \label{sec_performance}
The performance of Con-Pi framework is evaluated in a real environment and is compared with existing frameworks such as FogBus~\cite{FogBus} and FogPlan~\cite{FogPlan}. These frameworks are selected since they are open-source and easy to implement. In FogBus~\cite{FogBus}, the sensors forward the data to a master node. Later, the master node distributes the sensor data to multiple worker nodes where the applications are actually executed to process them. It is implemented using Java and PHP programming languages\footnote{https://github.com/Cloudslab/FogBus}. Conversely, in FogPlan, the data rate of IoT devices are tracked by a traffic monitor node. Based on the perception of this node, a Fog service controller node dynamically deploys applications on the other general-purpose Fog nodes to process the IoT data. It follows a Java-based implementation\footnote{https://github.com/ashkan-software/FogPlan-simulator}. For the experiments, the default topology of these frameworks (Con-Pi: mesh, FogBus: master-slave, FogPlan: hierarchical) are exploited as they are specified to be the best settings by their developers during individual evaluation. The experiment summary is given below.  
\begin{figure*}[!htbp]
  \centering
  \begin{minipage}[t]{0.3\textwidth}
    \centering
    \vspace{0pt}
    \includegraphics[width=\textwidth,align=t]{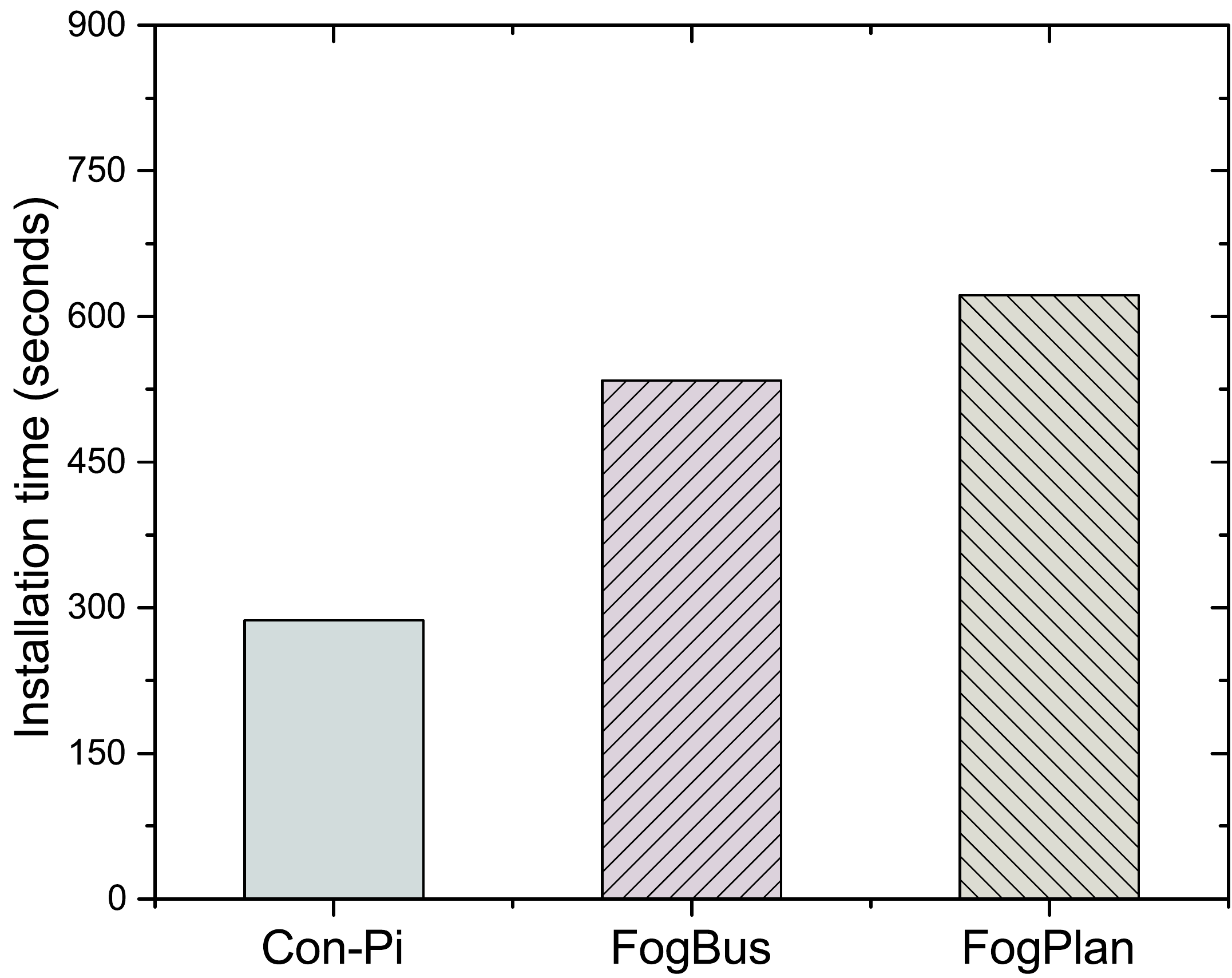}
    \caption{Installation time for different frameworks}
    \label{Fig:installation}
  \end{minipage}
  \qquad
  \hfill
  \begin{minipage}[t]{0.3\textwidth}
  \centering
    \vspace{0pt}
    \includegraphics[width=\textwidth,align=t]{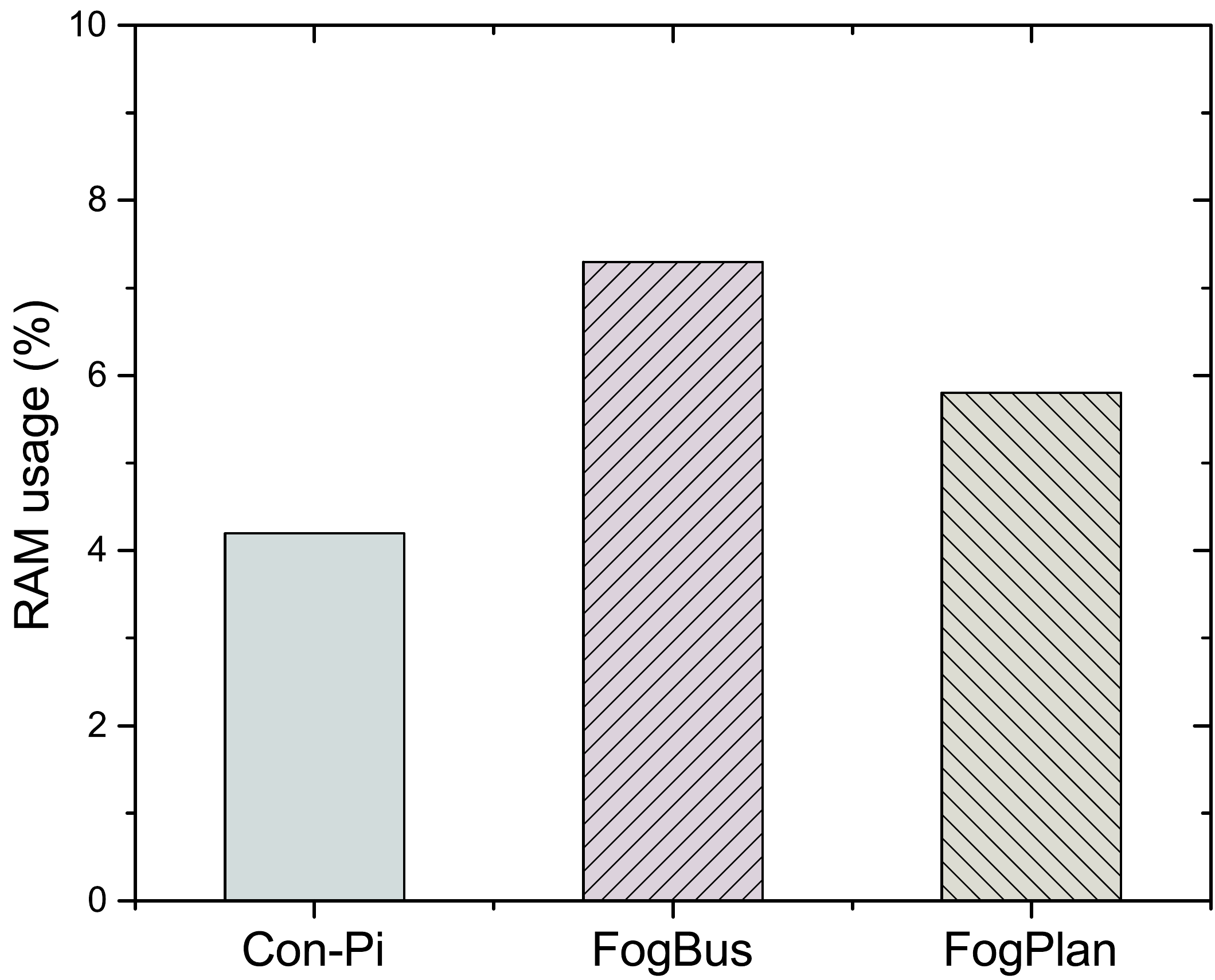}
    \caption{RAM Usage for different frameworks}
    \label{Fig:ram}
  \end{minipage}
  \qquad
  \hfill
  \begin{minipage}[t]{0.3\textwidth}
  \centering
  \vspace{0pt}
    \includegraphics[width=\textwidth,,align=t]{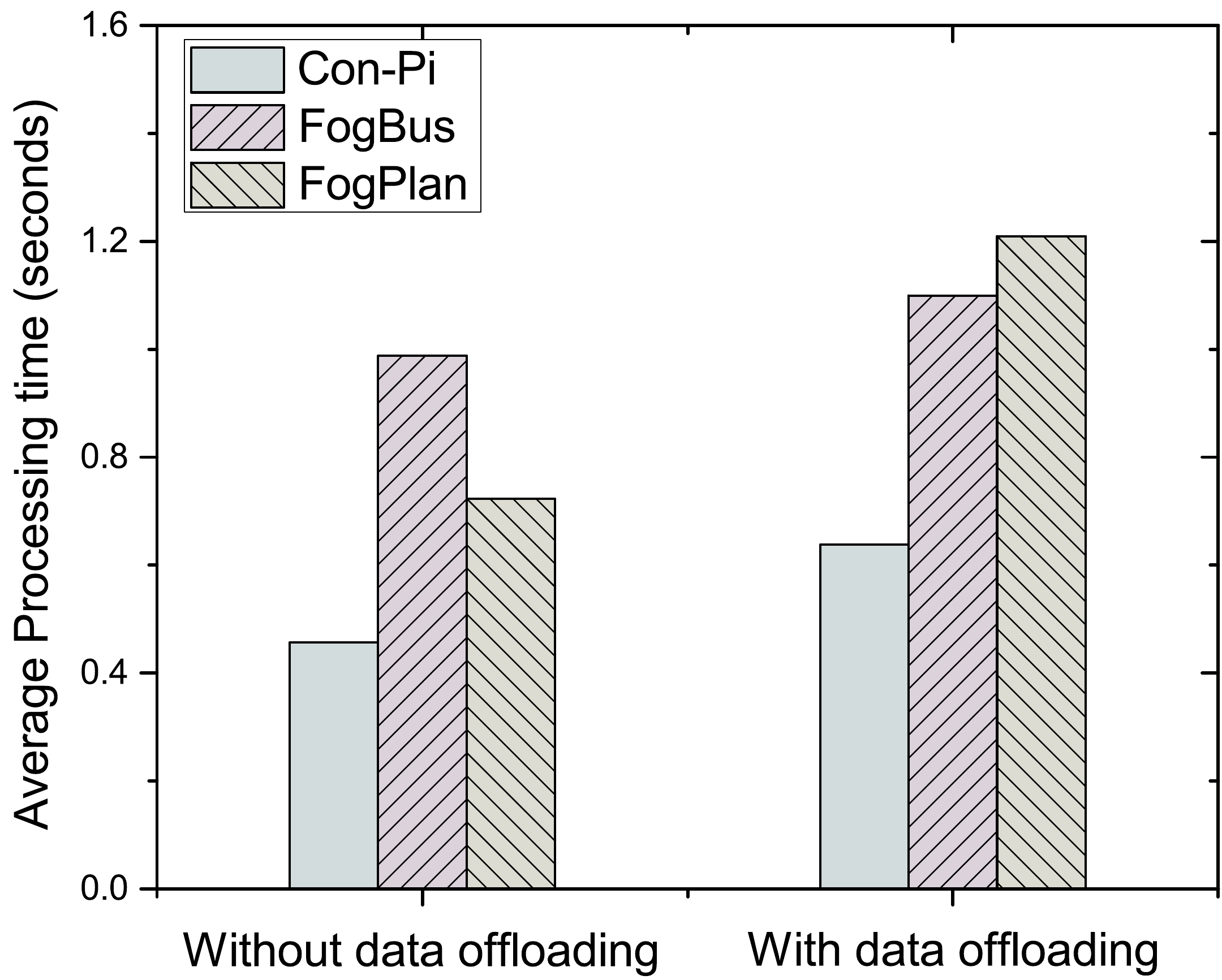}
    \caption{Input processing time on different frameworks}
    \label{Fig:processing}
  \end{minipage}
\end{figure*}
\subsection{Experimental Environment}
To model the experimental environment, six \textit{Raspberry PI 3 Model B+} devices have been used, and their uplink and downlink bandwidth are set to a specific value using \textit{WonderShaper} tool\footnote{https://github.com/magnific0/wondershaper}. For Con-Pi implementation, the RPis are organized in a mesh topology (as shown in Fig. \ref{Fig:Frame}). Conversely, for FogBus~\cite{FogBus}, one RPi is set as the master RPi, and the other three RPis are configured as the workers. On the other hand, for FogPlan~\cite{FogPlan}, two RPis are deployed as the Fog service controller and traffic monitor separately, and two RPis are set as the general-purpose Fog nodes. In both FogBus and FogPlan implementations, two RPis are solely used for input data generation for the corresponding application. Moreover, the Con-Pi, FogBus, and FogPlan frameworks individually enable the experiment environment to execute a deep neural networking (MobileNet-SSD\footnote{https://github.com/chuanqi305/MobileNet-SSD})-based image processing application. The application detects a specific set of objects from any given image\footnote{https://github.com/Redowan-Mahmud/ObjectDetection-RPi}, and there exist both Python and Java versions of the application with OpenCV (Open-Source Computer Vision Library) dependencies. Additionally, the application can be run both as a monolithic application and as a microservice in non-virtualized and virtualized physical resources, respectively. To accelerate the experiment, we have used a data set of birds with 200 species named CUB-200 \footnote{https://www.kaggle.com/data/126169} and emulate that the camera sensors integrated with the RPis are generating imagery inputs for the application in different frequencies. Table \ref{Tab_realSystem} lists the specifications of the experiment environment. 
\begin{table}[!t] 
\caption{Specification of experimental setup}\label{Tab_realSystem} 
\centering
\scriptsize
\begin{tabular}{|p{4 cm}p{3 cm}|}
\hline
\textit{Configuration of PI} &\\\hdashline[1pt/1pt]
Processor & Broadcom BCM2837B0 \\
RAM & 1 GB  \\
Battery capacity & 1820mAh \\
Clock & 1.4 GHz \\
Uplink & 2 MBPS \\
Downlink & 2 MBPS \\\hline
\textit{Application features} &\\\hdashline[1pt/1pt]
Number of bird images & 2000\\
Resolution (Pixel) & 640x480 \\
Average size & 0.230 MB \\
Container size &  756 MB\\
Image forwarding frequency & 0.5-2 images/second \\\hline
\end{tabular} 
\end{table} 
\subsection{Experimental Results}
In the experiments, the installation time of the frameworks, the number of processed inputs, data processing time, energy, CPU and RAM usage are considered as the performance metrics. Various tools, including Unix timestamp, PiJuice APIs and Linux top command, are used to observe these parameters. As the experiments are conducted in a real setup without any variation in resource configuration, network conditions, application features and input data, the results show a deterministic trend during the repetition. The experiment results are discussed as follows. 
%
\subsubsection{Installation Time}
%
%
%
Con-Pi is lightweight, as it requires a minimal number of third-party libraries for initiation. Moreover, Con-Pi is supported with an installation script that facilities simultaneous deployment of PiService, PiController and microservice on an RPi. Since the software system of Con-Pi is highly autonomous, it does not require any external entity to configure the RPis and make them interconnected. Therefore, the installation time for Con-Pi is comparatively lower than that of other frameworks (as shown in Fig. \ref{Fig:installation}). Nevertheless, to operate FogBus, several heavyweight software components, including Apache web server and MySQL database server, are required. Moreover, in FogBus, the worker and master RPis need to be configured individually. Due to these issues, the total installation time increases for FogBus. There exists a similarity between FogBus and FogPlan in terms of dependency on third-party software components. Consequently, they show similar trends for installation time. However, FogPlan requires additional time to configure the traffic monitor for RPi that affects installation time by 16\%.   
%
\subsubsection{Percentage of RAM Usage}
%
%
%
%
%
%
Con-Pi is a Python language-based framework, whereas FogBus and FogPlan are developed in Java. Typically, when a Python program requires more memory for its execution, the Python Virtual Machine automatically enhances RAM allocation up to a certain extent allowed by the corresponding operating system. On the other hand, the Java Virtual Machine (JVM) consumes a fixed amount of memory prior to executing any Java program. As a consequence, there exists a limited scope to adjust the memory for a Java program according to the context of the execution. Moreover, due to static typing, all the libraries referenced by a Java program have to be loaded separately in the permanent generation space of the JVM~\cite{ProgrammingLanguage}. As a result, Con-Pi consumes less RAM than that of other frameworks (as shown in Fig. \ref{Fig:ram}). Nevertheless, FogBus performs additional database operations during its initiation. As a consequence, its RAM usage elevates compared to the FogPlan framework, where such operations are not observed. 
\begin{figure*}[!htbp]
  \centering
  \begin{minipage}[t]{0.3\textwidth}
    \centering
    \vspace{0pt}
    \includegraphics[width=\textwidth,align=t]{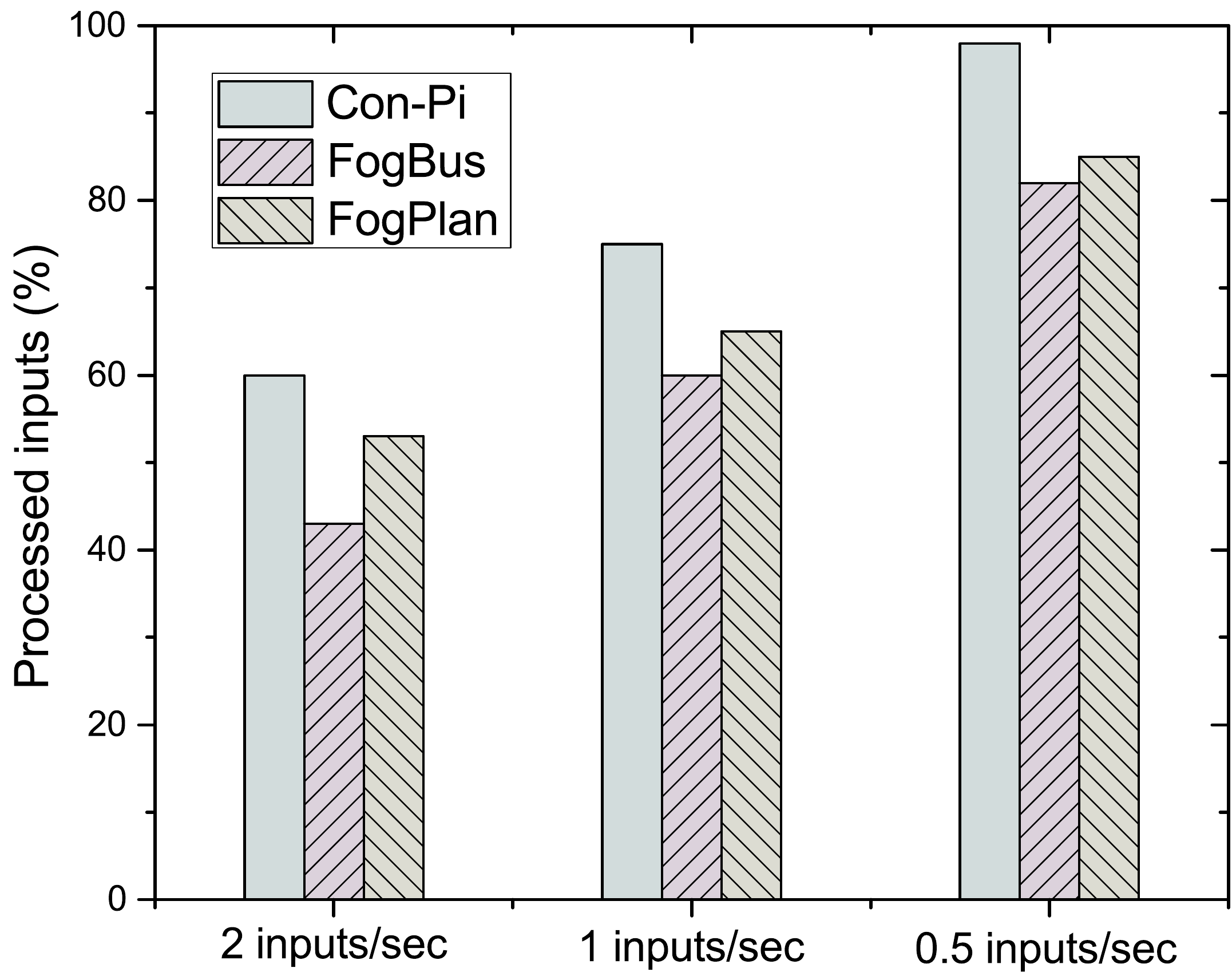}
    \caption{Percentage of processed input}
    \label{Fig:inputs}
  \end{minipage}
  \qquad
  \hfill
  \begin{minipage}[t]{0.3\textwidth}
  \centering
    \vspace{0pt}
    \includegraphics[width=\textwidth,align=t]{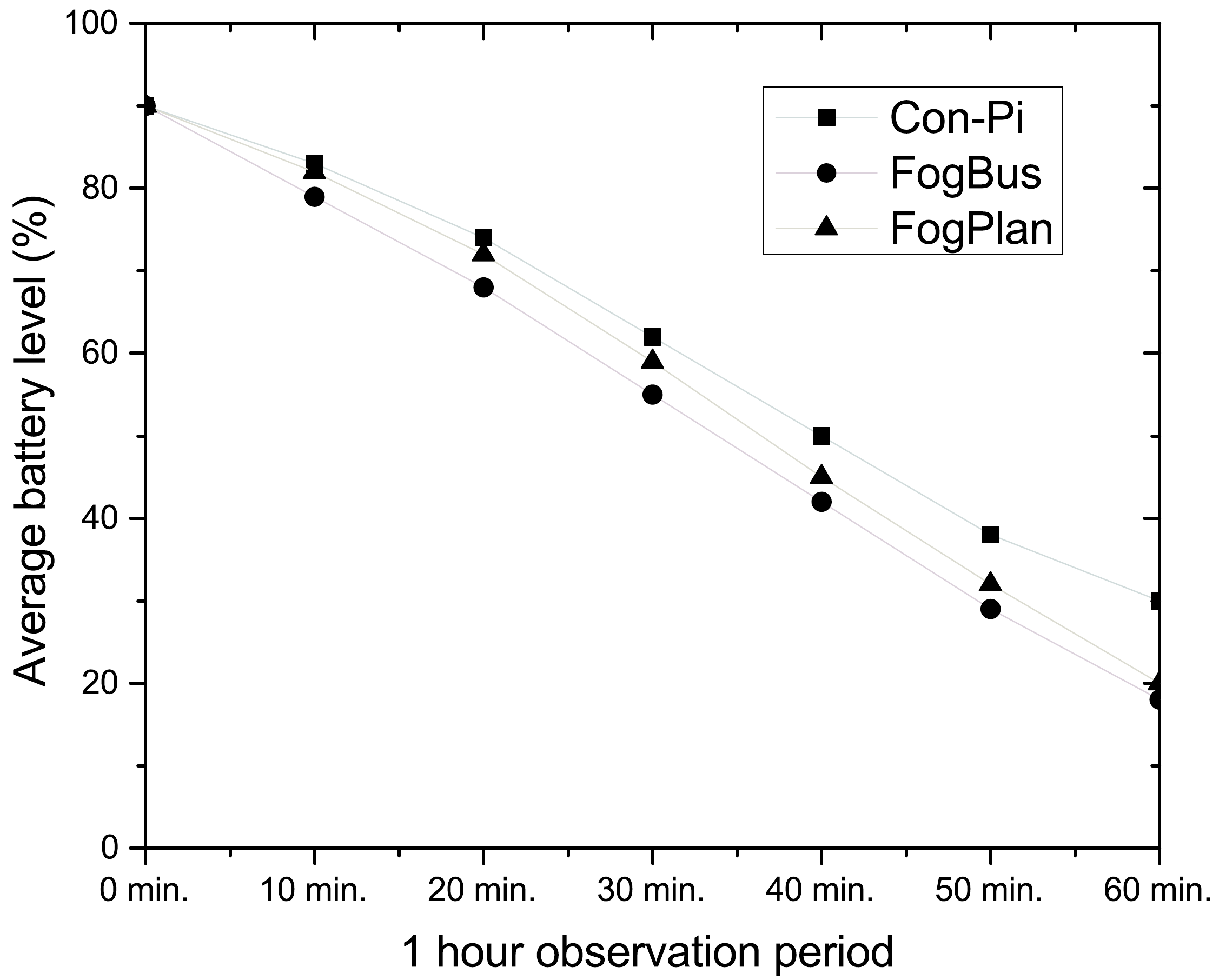}
    \caption{Battery level of RPis on different frameworks}
    \label{Fig:battery}
  \end{minipage}
  \qquad
  \hfill
  \begin{minipage}[t]{0.3\textwidth}
  \centering
  \vspace{0pt}
    \includegraphics[width=\textwidth,,align=t]{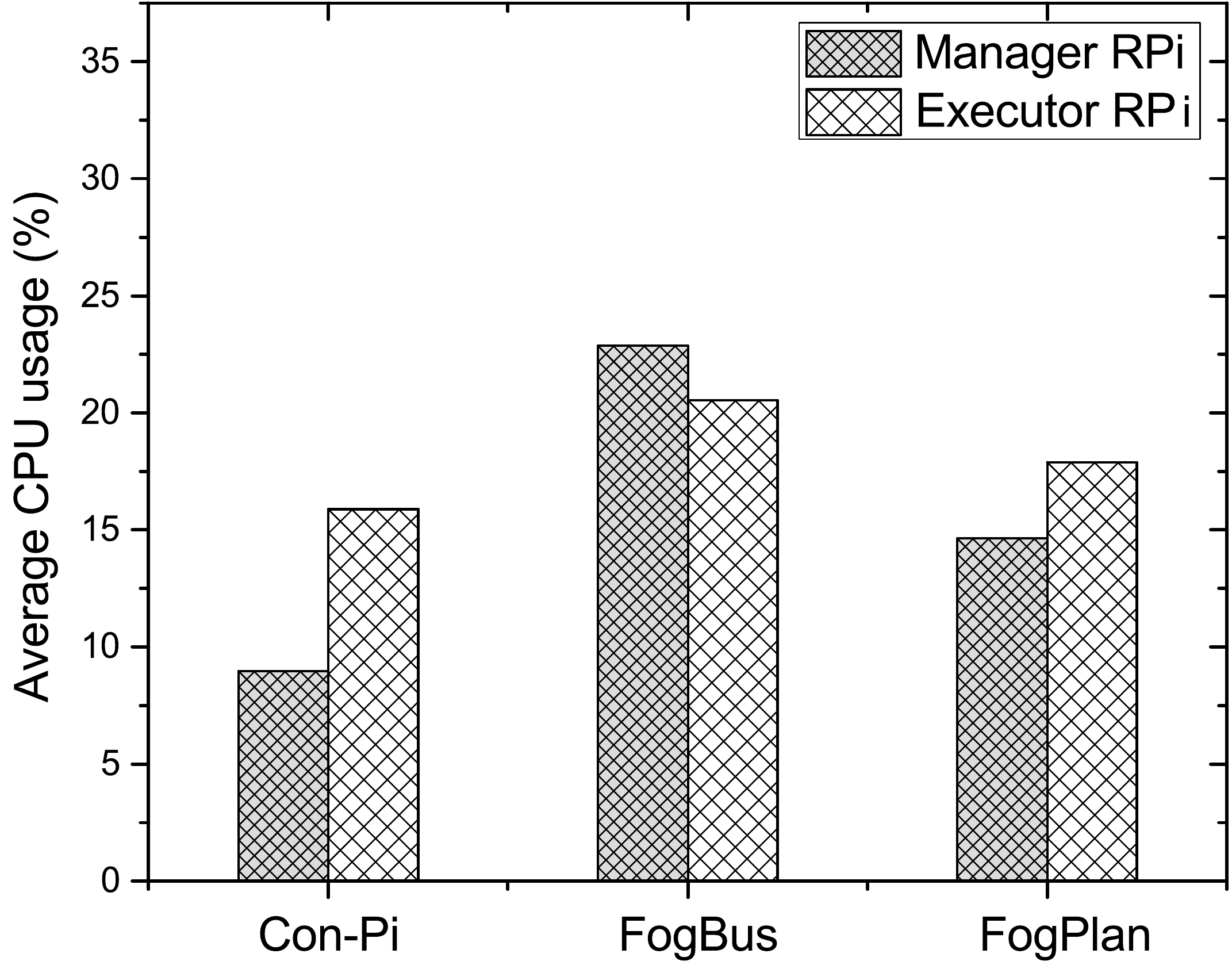}
    \caption{CPU usage of RPis on different frameworks}
    \label{Fig:cpu}
  \end{minipage}
\end{figure*}
%
%
%
%
%
%
\subsubsection{Data Processing Time}
This experiment has been conducted by switching the data offloading facility of Con-Pi framework. Inherently, Con-Pi supports the input data generation and processing on the same RPi. Due to lightweight service-oriented interactions and in-memory operation, Con-Pi can assist data processing within a shorter period when no offloading is occurred (as shown in Fig. \ref{Fig:processing}). FogPlan follows an almost similar principle to Con-Pi. Nevertheless, it consumes an additional time in processing the input data because of the multiple intermediate disk operations. On the other hand, while offloading data from one RPi to another, the average processing time unevenly increases for each frameworks due to their topological variations. However, the mesh topology of Con-Pi simplifies the handshaking between the data generator and the application executor. Therefore, only a one-hop distant data propagation delay adds to the actual processing time. Conversely, in FogPlan, multiple entities, including the data generator, traffic monitor, Fog service controller and general-purpose Fog nodes, work collaboratively in a hierarchical topology to realize the processing of offloaded data. Such interaction incurs a significant amount of communication delay that affects the total processing time of the input data in the FogPlan framework. Similarly, the input processing time elevates considerably in the master-slave topology of FogBus, where the data are frequently offloaded from the sensing node to the computing RPis via a broker (master node). 
%
%
\subsubsection{Percentage of Processed Inputs}
%
%
%
Fig. \ref{Fig:inputs} depicts the percentage of processed inputs on different data generation frequencies of the RPis. At a higher frequency, the percentage of processed inputs becomes lower on every framework. It happens because of the arrival of the next input data while processing the previous one. In such cases, the newly arrived input data remain unattended and discarded for the sake of the real timeliness of the system that eventually affects the percentage of processed inputs with respect to the amount of total generated inputs. However, in Con-Pi (without the offloading facility), this percentage is higher than that of FogBus and FogPlan as it inherently processes data quicker and allows the RPis to attend more inputs. Additionally, due to lesser communication overhead, FogPlan performs better than FogBus in this experiment. However, with the relaxation on data generation frequency, such performance difference gets minimized.
%
%
\subsubsection{Percentage of Battery Level}
To observe the impact of the frameworks on the battery lifetime of RPis, the average battery level of RPis within the experiment environment is set to 90\%. Later, the RPis are kept running for 1 hour, and the readings are accumulated periodically (every 10 minutes). Moreover, the resource management policy is statically set to start the data offloading when the battery lifetime of the RPis reaches 50\%. In such settings, it is observed that Con-Pi preserves the energy of RPis more rigorously than other frameworks, especially when the average battery lifetime becomes lower than 50\% (as shown in Fig. \ref{Fig:battery}). It happens because Con-Pi can significantly reduce its computing overhead on energy-constrained RPis by lessening the functionalities of PiController and microservice. Since the software system of Con-Pi is highly modular, such operations can be easily conducted without degrading or affecting the performances of Con-Pi frameworks in other RPis. However, this sort of facility is not available in FogBus and FogPlan. Moreover, the performance of FogBus degrades in this experiment over the course of time as there exists no option to save energy for the master RPi. Conversely, FogPlan maintains an event-driven interaction with the Fog service controller and traffic monitor, which saves energy for these nodes and eventually improves the average battery level of the RPis. 
%
\subsubsection{Percentage of CPU Usage}
While offloading data using a specific framework, one RPi explicitly works as the manager and the other RPis function as the executor. For example, in Con-Pi, the RPi discovering the offloading destination is the manager RPi and the destination RPis are the executor. Similarly, in FogPlan, the Fog service controller is the manager RPi, and the general-purpose Fog nodes can be the executor RPis. Additionally, in FogBus, the master RPi is the manager, and any worker RPi can play the role of the executor. Fig. \ref{Fig:cpu} depicts the distribution of load among the manager and executor RPis. The average CPU usage of executor RPis does not vary significantly for both Con-Pi and FogPlan frameworks. However, while dealing with the equal number of executor RPis (three in this experiment), the CPU usage of manager RPi for Con-Pi is lower compared to FogPlan. It happens because the manager in Con-Pi only forwards data after making the handshaking with the executor RPis, whereas in FogPlan, the manager RPi needs to process the outcome of traffic monitor consistently. On the other hand, FogBus requires synthesising multiple programming languages such as PHP (web service) and Java at both manager and executor RPis, which elevates the CPU usage more than other frameworks. 
\section{A Case Study} \label{sec_case} 

\begin{figure*}[!htp]
  \centering
    \centering
    \includegraphics[width=1\textwidth]{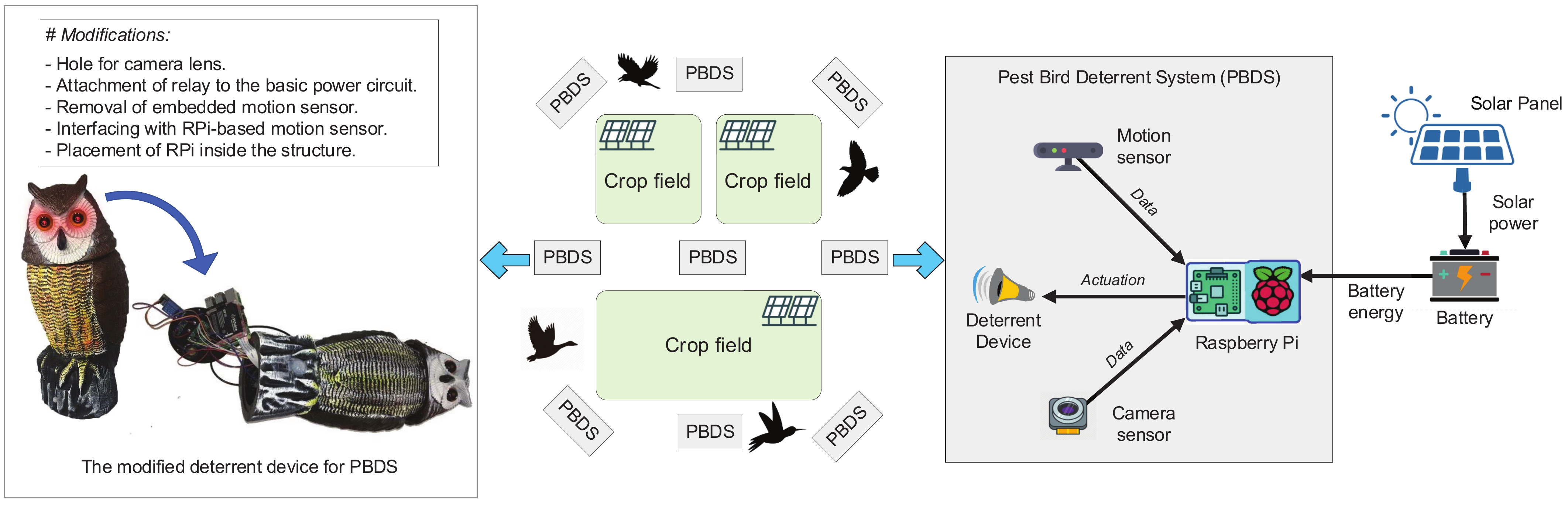}
	\caption{Con-Pi based Pest Bird Deterrent Systems}
	\label{Fig:prototype}
\end{figure*}

Con-pi enables developers to create and deploy their applications in the form of microservices across edge nodes without worrying about the underlying infrastructure and complexities of resource allocations. Thus, it can be adopted by various IoT-edge use cases. In particular, due to its fully distributed control plane and its built-in support for managing renewable energy sources, it can be used for smart applications in remote areas. In such scenarios, there is either no network coverage or a network with limited bandwidth. In addition, access to grid electricity is difficult, if not impossible. Examples of such applications include but are not limited to unmanned agricultural machinery, an early warning system for fires in frosts, and oil drilling. In the following, we will discuss the pest bird deterrent application as a use case scenario for Con-Pi.

The monetary loss for pest birds in Australia is roughly AUD \$313 million, which is the highest among all vertebrate pests~\cite{Australia}. Often, pest birds are speedy and do significant damage to crops quickly. Therefore, to detect and deter them from the farming area, the sense-analysis-actuation loop must be executed in real-time. In remote and vast farming areas, the arrangement of electricity to run a bird deterrent system is also tedious and costly. To address these challenges, we have developed a prototype of an automated Pest Bird Deterrent System (PBDS). The PBDS detects and identifies pest birds and deters them by frightening them. The Con-Pi framework has been adopted for deploying and operating the PBDS. As noted, Con-Pi enables Edge/Fog platforms on low-cost RPis and incorporates in-built supports for resource sharing and task offloading, which are highly required for the proposed PBDS. Con-pi also ensures efficient container management that helps energy harvesting and increases the lifetime of RPis within the PBDSs.
\begin{figure}[!htp]
\centering 
\includegraphics[width=1\columnwidth]{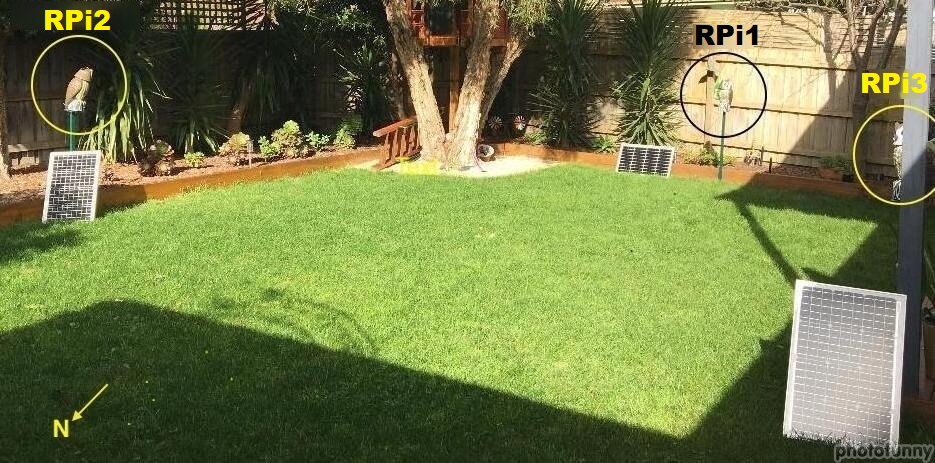}
\caption{The test PBDS setup for a trial in an urban backyard (Photo shoot at 13:12 on July 13, 2020).}
\label{Fig:backyard}
\end{figure}
\begin{figure}[!htp]
\centering 
\includegraphics[width=0.50\columnwidth]{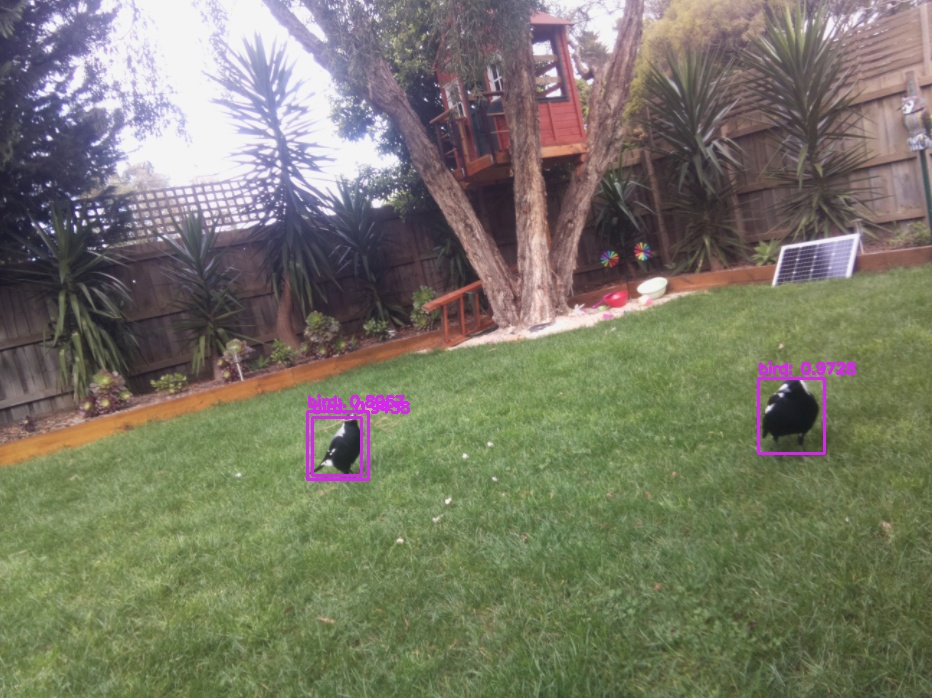}
\caption{A sample output image from the backyard trial.}
\label{Fig:birds}
\end{figure}

\par The developed PBDSs are deployed in multiple places of the farming area, as it is shown in Fig.~\ref{Fig:prototype}. The hardware components of a PBDS includes a motion sensor that tracks any movement within the farming areas and activates a camera sensor for capturing the images. Data from these sensors are directly forwarded to an RPi for processing through an application. A sample deterrent device that produces sounds, makes movement, emits laser beams and is connected with the RPi through the relay is also shown in Fig.~\ref{Fig:prototype}. The fake predatory bird device used as a deterrent has been structurally modified so that it can contain all other hardware components, including the RPi inside itself. Furthermore, we used solar panels and batteries to operate the RPi of PBDSs. As a sample case study, the solar panels are deployed in an urban backyard of the greater Melbourne area (Fig. \ref{Fig:backyard}), and their absorbed solar power is accumulated in the batteries. Later, the batteries dispatch the energy to the RPIs. Within the RPi, the data processing application is executed on a Docker container and helps in detecting the pest birds. We have used SSD (Single Shot MultiBox Detector) library, a state-of-the-art real-time object detection system, and OpenCV to perform pest bird detection. A sample image output of the PBDS with detected birds is shown in Fig.~\ref{Fig:birds}. Upon identifying the pest birds, the application actuates the relay to switch on the deterrent device to deter the pest birds\footnote{https://www.youtube.com/watch?v=-4T0hzdZaQM}. 
\begin{table}[!t]
\footnotesize 
\centering
\caption{System configurations}\label{tab_config}
\vspace{-0.25cm}
\begin{tabular}{|p{2.3cm}p{5.7cm}|}
\hline
    PBDS components & Description \\ \hline
    RPis & Raspberry Pi 3 Model B+, Broadcom BCM2837B0, Cortex-A53 (ARMv8) 64-bit SoC, 1.4GHz, 1GB LPDDR2 SDRAM, IEEE 802.11.b/g/n/ac wireless LAN, Bluetooth 4.2, BLE, Extended 40-pin GPIO.\\ \hline 
    PiJuice HAT & BP7X Motorola Droid 2 (A955), 1820mAh, 3.7V Lithium-Ion battery.\\ \hline
    Motion sensor & HC-SR501 Passive Infrared (PIR) Motion Sensor, Logic output 3.3V/0V.\\ \hline
    Camera sensor & RPi NoIR Camera Board v2, 8 megapixel Sony IMX219, Optical size of 1/4 inch.\\ \hline
    Solar panel & PiJuice solar panel, 40Watt, 5V/35Watts regulated maximum output.\\ \hline
    Relay & JQC-3F, 5V Single Channel Relay 10 Amp.\\ \hline
    Deterrent device & Whites group solar power owl, Model 18404.\\ \hline
\end{tabular}
\end{table}

\begin{figure*}[!t]
\centering
\subfigure[Battery level]{\includegraphics[width=0.33\textwidth]{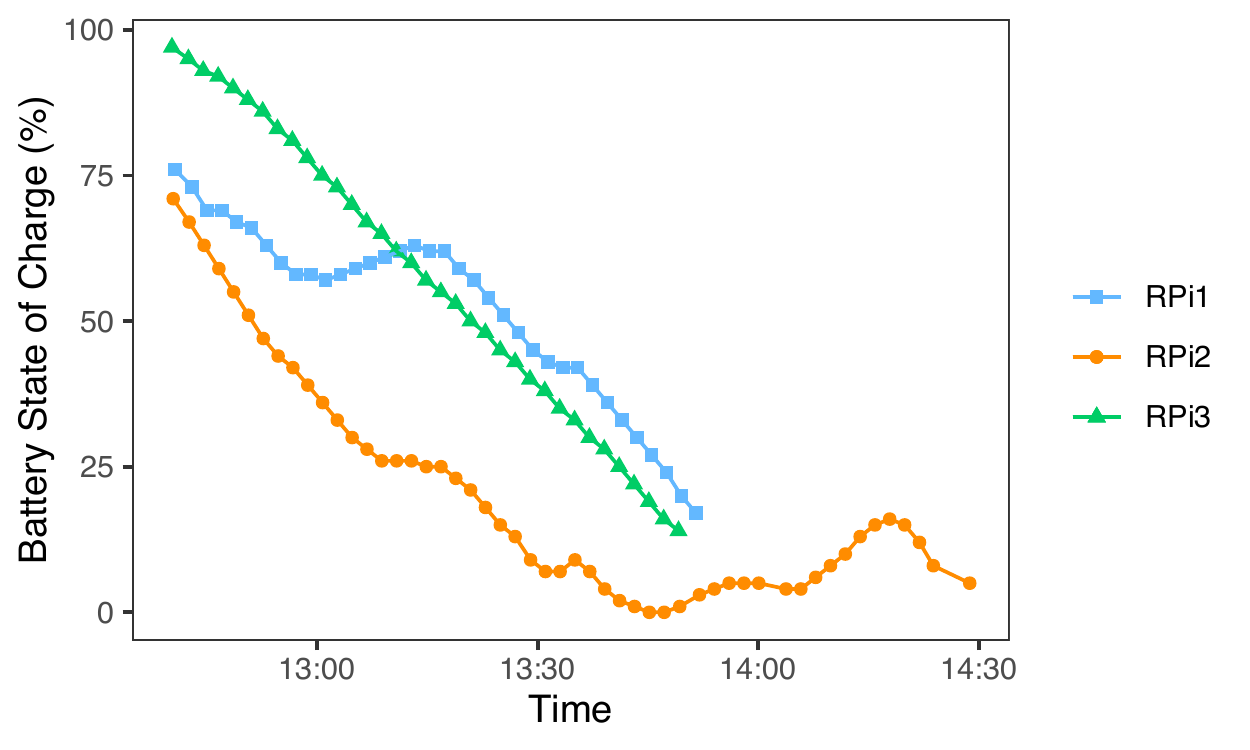}\label{Fig:batteryPi}}
\subfigure[CPU utilization]{\includegraphics[width=0.33\textwidth]{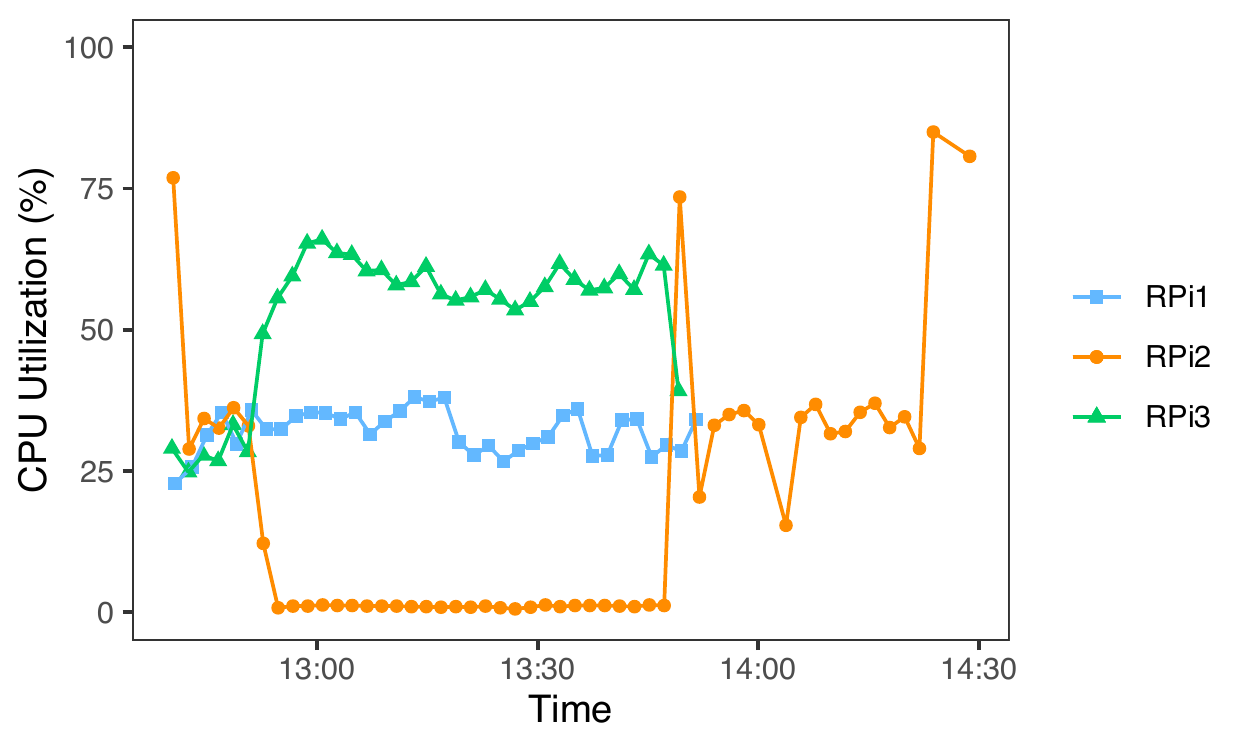}\label{Fig:CPU-Util}}
\subfigure[Number of Containers]{\includegraphics[width=0.33\textwidth]{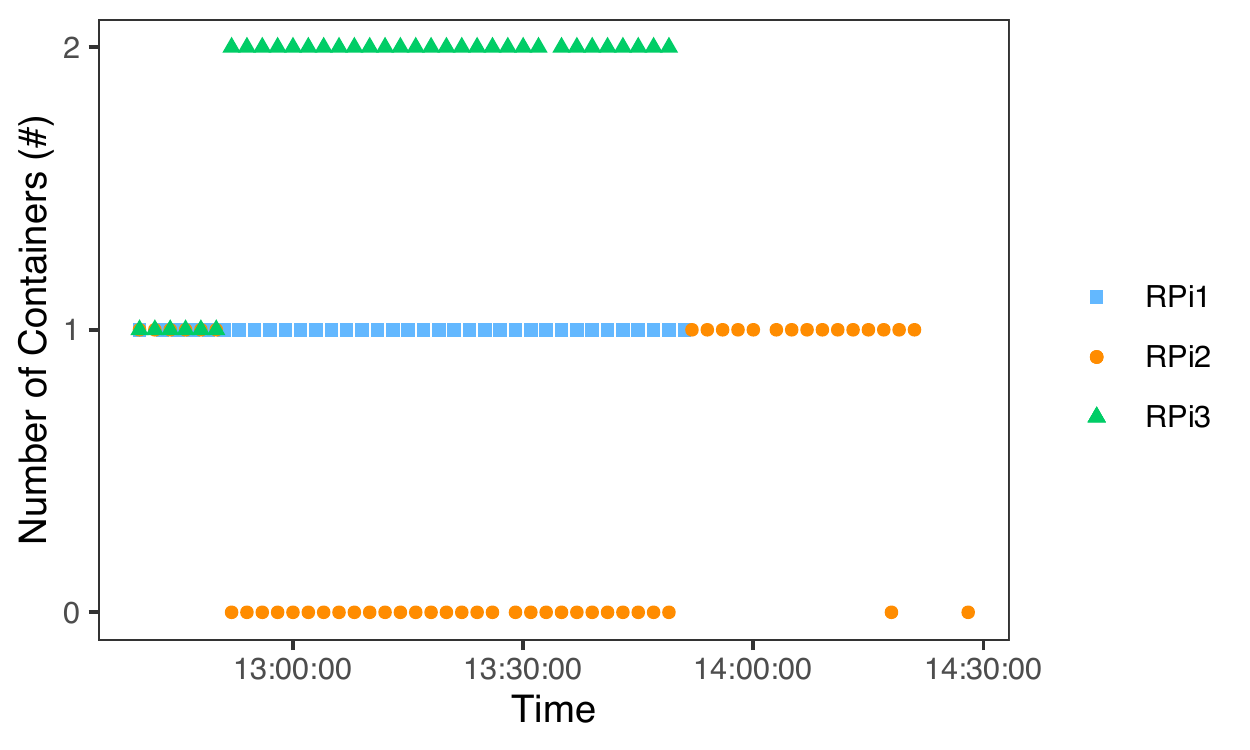}\label{Fig:Con}}
\caption{The Battery State of Charge, CPU Utilization, and Number of  Containers for RPis on Con-Pi in different time of the day (12:30 - 14:30)}
\label{Fig:test}
\end{figure*}
\par The Con-Pi framework automatically discovers PBDS devices and initiates task (container) offloading when the battery charge level of hosting RPis become low. Since PBDSs are connected through a dedicated local network, the internal exchanges of small-size images and actuation commands are performed in an insignificant timespan. 
\par To investigate the efficacy of the Con-Pi framework with respect to the case study, we conducted a 2-hour experiment on an experimental setup depicted in Fig. \ref{Fig:backyard}. The configurations of hardware components within the PBDSs are listed in Table \ref{tab_config}. The experiment was dated July 13, 2020, and the weather condition was partly cloudy; therefore, solar radiation intensity was up and down during the test. We used the proposed policy in Section~\ref{Sec:policy} for energy management. The values for $\alpha$, $\beta$ and $\gamma$, in Algorithm~\ref{algo-share}, are set as 50\%, 60\% and 70\%, respectively.
\par Results show that Con-Pi successfully operates the PBDS and effectively performs resource sharing and task offloading for energy conservation on battery-powered RPis. Fig.~\ref{Fig:test}  depicts the battery state of charge, CPU utilization, and the number of Docker containers for each of the three RPis in the experiment. At the beginning of the test, RPi3 has the highest level of battery charge, followed by RPi1, and RPi2 (Fig.~\ref{Fig:batteryPi}). Gradually, as the battery level at RPis decreases, RPi2, which has the lowest battery charge, reaches 50\% charge before other devices ($\alpha=50\%$), where at 12:52 it offloads its Docker container to RPi3 with the battery charge over 70\% (Fig.~\ref{Fig:Con}). By the time RPi2 and RPi1 battery levels go below 50\%, there are no other devices in the system with the battery level higher than 70\%. Therefore, no other task offloading occurs in the system. By 13:49, RPi3 runs out of charge and is switched off, RPi2 starts its container locally again (Fig.~\ref{Fig:Con}). Note that the sudden high CPU utilization on RPi2 at this time happens due to boot-up process of a new container. At around 14:00, the battery state of charge of RPi2 is also increasing while the CPU utilization is higher than before. This is an indication of higher solar irradiation at that time. The same observation can be made during 13:00 to 13:1 for RPi1. The offloading process allows RPi2 to stay in working condition along with other RPis up to 13:49. It is worth mentioning that the battery level of RPi2 reduces faster than others at the beginning of the experiments before 12:52 while all RPis running one container. This happened because RPi2 is in the shade of a tree at this period, and energy generation by the solar panel is lower compared to the other two devices receiving full sun exposure. The task offloading feature provided by Con-Pi allows RPi2 to remain operational during the low energy generation period and slow down the battery drain preventing a potential early flat battery compared to other RPis.

\par Moreover, as it is shown in Fig.~\ref{Fig:CPU-Util}, the CPU utilization of RPi devices complies with the container placements in the system. Each Docker container hosting a pest bird deterrent application on average utilizes the CPU by roughly 30\%. As shown in Fig.~\ref{Fig:CPU-Util}, between 12:52 and 13:49, when RPi2 does not run any Docker containers, the CPU utilization of RPi2 is nearly 0\%. While Con-Pi uses network resources (Wi-Fi) to transfer sensory data (images and motion data) to the offloaded container during this period, its PiController and PiService modules barely use CPU resources. This overall energy saving shows the potential of Con-Pi in building smart CPSs on resource constraint devices for IoT and meets the objective of energy-aware resource management discussed in Section \ref{RMS}.
\section{Future Works} \label{sec_future}
The software components of Con-Pi are extensible. Hence, there exists a significant scope to incorporate extra features into the existing Con-Pi framework. Some of the relevant research directions are listed below. 
\par $\bullet$ \textit{Lightweight security feature}: Through containerization, Con-Pi facilitates explicit isolation of the applications. However, it is also required to secure the resources of the RPi while dealing with the external entities. Currently, Con-Pi assumes that all RPis within a computing environment are reliable and exclusively managed by the system operators, which is not applicable in all cases. Therefore, efficient security measures need to be added to the Con-Pi framework. At the same time, such measures should be lightweight so that they do not degrade performance significantly. 

\par $\bullet$ \textit{Resource management policies}: As noted, Con-Pi supports the integration of customized resource management policies. However, Con-Pi's built-in resource discovery and resource sharing policies are purposely generic (targets homogeneous microservices) and capable of dealing with one-dimensional decision-making parameters such as energy or response time. Given the highly dynamic nature of Fog and Edge computing environments, different and more complex resource management policies with multiple driving factors must be integrated with the Con-Pi framework. 
Furthermore, the support for LoRaWAN and Bluetooth can also be integrated with Con-Pi besides WiFi to enable it for energy-efficient communications. 

\par $\bullet$ \textit{Serverless Architecture}: We aim to extend Con-Pi via the adoption of Serverless architecture. The Serverless architecture will allow application developers to only focus on the development of their application, and Con-Pi will take care of running applications, provisioning resources and scaling them with high availability. To adopt the Serverless architecture, we rely on the event-based nature of IoT applications and container-based architecture of Con-Pi. This is viable for Con-Pi as its software system is wholly modulated, and applications can be seen as functions in the Function-as-a-Service model provided by Serverless computing.
\section{Conclusions} \label{sec_conclusions}
%
In this work, we developed Con-Pi, a framework to realize these two paradigms by harnessing the computing capabilities of Raspberry Pis (RPis). Con-Pi facilitates containerization of Internet-of-Things applications using Docker Containers so that they can be executed in the form of lightweight microservices. It also facilities seamless integration of customized resource sharing, task offloading and resource discovery policies to manage the computing environment. Additionally, its performance in terms of data processing time, installation time, CPU and RAM usage are compared with contemporary frameworks. The comparison outcomes demonstrate the efficiency of the proposed framework to a great extent. Moreover, a pest bird deterrent system has been developed using Con-Pi that helps to demonstrate its real-world applicability. The potential future directions highlighted in this work can also contribute to enhancing the feasibility of the proposed Con-Pi framework.
\section*{Software Availability} 
The current version of Con-Pi has been launched as an open-source. Its executable scripts, dependency libraries and candidate microservice image can be found at \textit{https://bitbucket.org/disnet-lab/con-pi/}

\section*{Acknowledgements}
We would like to thank M. Aslanpour for his help in improving our paper. We also thank Monash Animal Ethics Team for approving the animal ethics application related to the case study. We are indebted to Dr. R. Clarke and B. Viola for helping us to prepare the animal ethics application. This project was supported by AgTech Priming Grant Program funded by Monash University in collaboration with Bosch.
\bibliographystyle{IEEEtran}
\bibliography{Con-Pi}

\vskip -1\baselineskip plus -1fil
\begin{IEEEbiography}[{\includegraphics[width=1in,height=1.25in,clip,keepaspectratio]{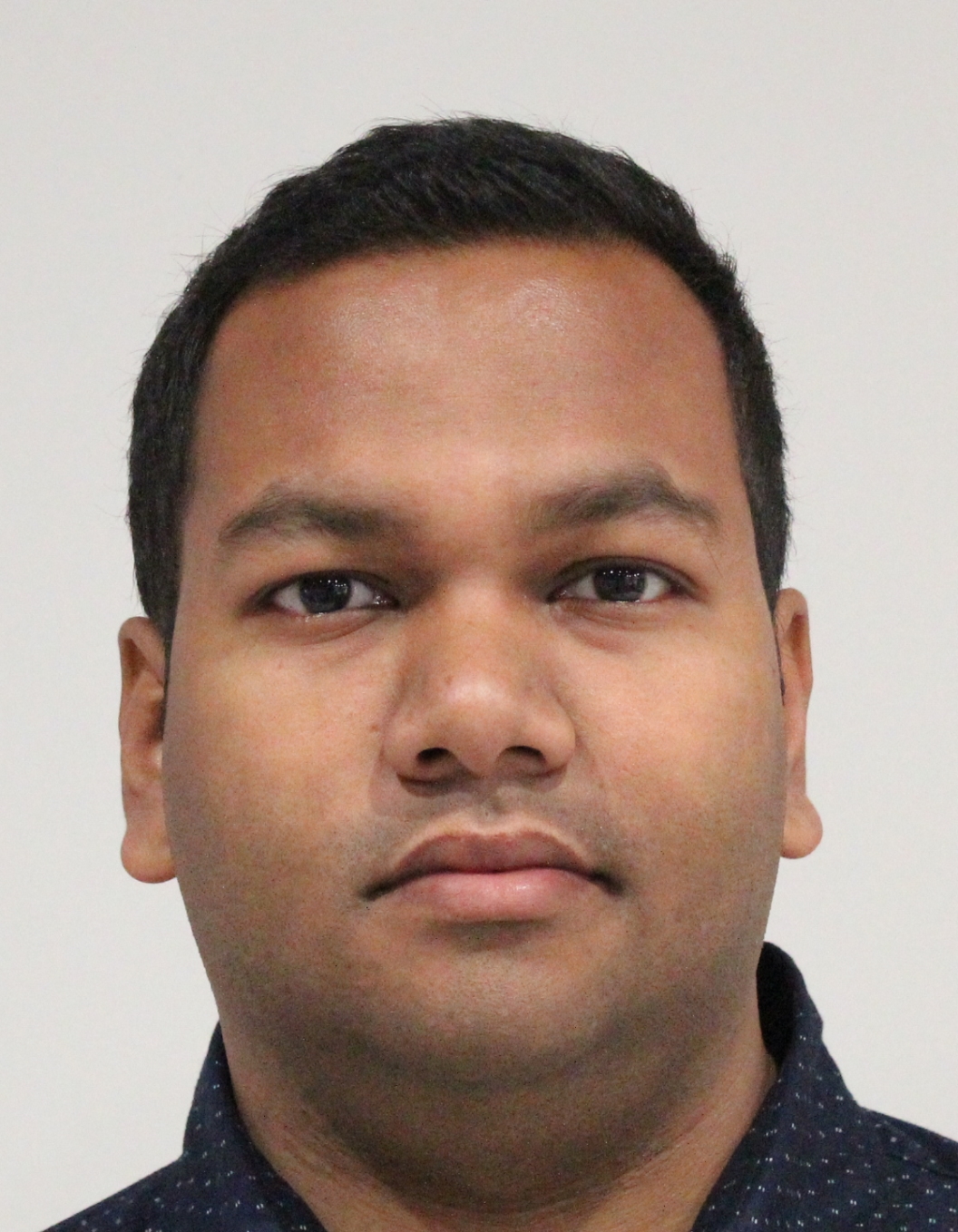}}]{Redowan Mahmud} is currently working as a Research Assistant in the Faculty of Information Technology, Monash University. He received Ph.D. in 2020 from the School of Computing and Information Systems, the University of Melbourne, Australia. He received B.Sc. degree in 2015 from Department of Computer Science and Engineering, University of Dhaka, Bangladesh. His research interests include Internet of Things, Fog and Edge computing. 
\end{IEEEbiography}
\vskip -0.5 \baselineskip plus -1fil 
\begin{IEEEbiography}[{\includegraphics[width=1in,height=1.3in,clip,keepaspectratio]{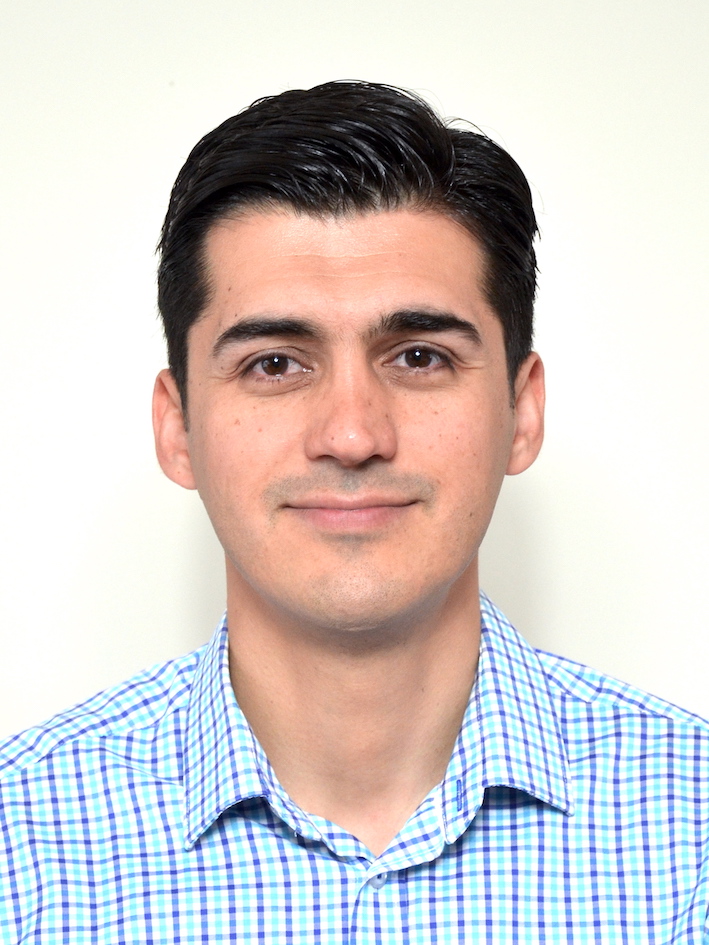}}]{Adel N. Toosi} is a lecturer (a.k.a. Assistant Professor) in Computer Systems at Department of Software Systems and Cybersecurity, Faculty of Information Technology, Monash University, Australia. Before joining Monash, Dr Toosi was a Postdoctoral Research Fellow at the University of Melbourne from 2015 to 2018. He received his PhD degree in 2015 from the School of Computing and Information Systems at the University of Melbourne. Dr Toosi’s research interests include scheduling and resource provisioning in Cloud/Fog/Edge Computing, Internet of Things, Software-Defined Networking, Green Computing environments. Currently, he is working on green energy management for Edge/Fog computing environments. For further information, please visit his homepage: http://adelnadjarantoosi.info.
\end{IEEEbiography}

%








\end{document}